\documentclass{article}
\usepackage{arxiv}

\usepackage[utf8]{inputenc} 
\usepackage[T1]{fontenc}    
\usepackage{hyperref}       
\usepackage{url}            
\usepackage{booktabs}       
\usepackage{amsfonts}       
\usepackage{nicefrac}       
\usepackage{microtype}      
\usepackage{lipsum}
\usepackage{framed,multirow}
\usepackage{rotating}
\usepackage{tabularx}
\usepackage{makecell}
\usepackage{amssymb}
\usepackage{latexsym}
\usepackage{caption}
\usepackage{subcaption}
\usepackage{xtab}
\usepackage{longtable}
\usepackage{afterpage}
\newcolumntype{Y}{>{\centering\arraybackslash}X}
\newcolumntype{S}{>{\hsize=.35\hsize}X}
\newcolumntype{T}{>{\hsize=.25\hsize}X}
\graphicspath{ {./images/} }

\usepackage{xcolor}
\usepackage{pdflscape}

\usepackage[backend=bibtex,natbib=true,maxcitenames=1,maxbibnames=5,giveninits=true]{biblatex}
\usepackage{graphicx}
\usepackage{tikz}
\usepackage{forest}
\usetikzlibrary{trees,positioning,shapes,shadows,arrows.meta,chains}
\usepackage{pgfplots}
\pgfplotsset{compat=1.17} 
\usepackage{pgf-pie}

\newcommand{\numpapers}{171}

\title{The HoloLens in Medicine: A systematic Review and Taxonomy}

\author{
  Christina Gsaxner\\
  Institute of Computer Graphics and Vision\\
  Graz University of Technology\\
  8010 Graz, Austria \\
  \texttt{gsaxner@tugraz.at} \\
  \And
  Jianning Li \\
  Institute of AI in Medicine\\
  University Medicine Essen\\
  45131 Essen, Germany \\
  \And
  Antonio Pepe \\
  Institute of Computer Graphics and Vision\\
  Graz University of Technology\\
  8010 Graz, Austria \\
  \And
  Yuan Jin \\
  Institute of Computer Graphics and Vision\\
  Graz University of Technology\\
  8010 Graz, Austria \\
  \And
  Jens Kleesiek \\
  Institute of AI in Medicine\\
  University Medicine Essen\\
  45131 Essen, Germany \\
  \And
  Dieter Schmalstieg \\
  Institute of Computer Graphics and Vision\\
  Graz University of Technology\\
  8010 Graz, Austria \\
  \And
  Jan Egger \\
  Institute of AI in Medicine\\
  University Medicine Essen\\
  45131 Essen, Germany \\
}

\begin{document}
\maketitle
\begin{abstract}
The HoloLens (Microsoft Corp., Redmond, WA), a head-worn, optically see-through augmented reality display, is the main player in the recent boost in medical augmented reality research. In medical settings, the HoloLens enables the physician to obtain immediate insight into patient information, directly overlaid with their view of the clinical scenario, the medical student to gain a better understanding of complex anatomies or procedures, and even the patient to execute therapeutic tasks with improved, immersive guidance. In this systematic review, we provide a comprehensive overview of the usage of the first-generation HoloLens within the medical domain, from its release in March 2016, until the year of 2021, were attention is shifting towards it's successor, the HoloLens 2. We identified \numpapers~relevant publications through a systematic search of the PubMed and Scopus databases. We analyze these publications in regard to their intended use case, technical methodology for registration and tracking, data sources, visualization as well as validation and evaluation. We find that, although the feasibility of using the HoloLens in various medical scenarios has been shown, increased efforts in the areas of precision, reliability, usability, workflow and perception are necessary to establish AR in clinical practice. 
\end{abstract}


\twocolumn

\section{Introduction} 

Augmented Reality (AR) enhances the real world with virtual content by interposing computer graphics between the human eye and its field of vision. Recent consumer-oriented developments made AR devices accessible to the general public. As a result, the AR field saw a strong growth in various domains, such as industry and entertainment. A main player in this new development was the HoloLens (Microsoft Corp., Redmond, WA), released in 2016. The HoloLens was originally marketed for applications in gaming, communication and 3D modeling; nevertheless, it quickly drew the attention from the medical domain. This development is unsurprising -- after all, one can hardly imagine a professional domain in which AR could have a more significant impact than in medicine. AR has the potential to grant physicians "X-Ray vision" -- the ability to see critical structures within the patient, without making a single incision. Wearable devices, such as the HoloLens, can make critical patient information permanently and readily available, and show them directly in the vision of the physicians. This approach allows them to keep their focus on the patient only and makes surgeries cluttered with monitors obsolete. Immersing remote experts into the mixed reality environment would further permit more and more patients to benefit from their expertise. Patients could be monitored and guided through various treatment and rehabilitation stages using AR, be it within the clinic or in their own homes, while medical students could practice critical interventions in a safe, virtually enhanced setting or immerse themselves in 3D anatomy. The HoloLens 1, as the first wearable, fully untethered AR device, was certainly an important step towards the future of AR in medicine. But how much could it contribute, and how far are we in making the aforementioned scenarios a reality?

In this systematic review, we provide a comprehensive overview of works that reported the usage of the first-generation HoloLens in the medical domain from 2016 to 2021. We identified \numpapers~relevant publications through a systematic search of the PubMed and Scopus databases and analyzed them according to their intended use case, technical methodology concerning registration and tracking, data sources and data visualization, as well as evaluation and validation. Throughout our review, we highlight principal findings, identify gaps and discuss challenges and limitations. With the recent availability of its successor, the HoloLens 2, this review outlines the impact the first generation HoloLens had during its lifetime in the medical area. 


\section{Background}
\subsection{Augmented reality}
One of the most common definitions of AR stems from the virtuality continuum definition by Milgram et al.~\cite{milgram1994taxonomy}, who describe AR as a mixed reality (MR), which contains mainly real elements, enhanced with virtual content. Azuma et al.~\cite{azuma1997survey} further characterize AR environments as combining reality and virtuality by registration in 3D, while being interactive in real-time. Although this definition makes clear that AR can appeal to all senses, it is mostly concerned with visual data. In the medical field, where digital imaging techniques provide rich information, AR has huge potential. Unsurprisingly, once technology was advanced enough to consider real-world AR applications, it quickly drew the attention from the medical domain. 

\paragraph{AR displays.}
The first medical AR systems were introduced as early as the late 1980s, with Roberts et al.~\cite{roberts1986frameless} describing the first system, an operating microscope augmented with segmented computed tomography (CT) images. A head-mounted displays (HMD) continued to be a popular display choice in early medical AR systems, as, for example, demonstrated by the works in the 1990s~\cite{state1996technologies, fuchs1996towards} and in the early 2000~\cite{sauer2001head}, who developed a video see-through HMD for medical applications. An HMD is a natural choice for medicine, as they intuitively align the head gaze of the wearer with the viewpoint of the content, and keep the hands of the wearer free at the same time. However, early HMD designs could not easily fulfill the high demands of medical AR systems in terms of performance, latency and accuracy, which require powerful computational infrastructure. Usually, this challenge resulted in bulky form factors, with wired connections between HMD and more capable computing and tracking infrastructure, making these systems difficult to implement in real clinical scenarios. Although head-worn microscopes, optical and video see-through displays continued to be relevant, in the years between 2011 and 2017, we see a shift towards world-localized displays, such as stationary monitors or projector systems~\cite{eckert2019augmented}. The release of the HoloLens 1, which was the first self-contained AR-HMD with a slim form factor, subsequently caused research attention to shift towards optical see-through (OST) displays again~\cite{gsaxner2021archapter}.

\paragraph{Registration and tracking.}
Alignment between reality and virtuality is a fundamental concept of AR, which is realized via registration. In a medical context, registration is mostly desired between medical data, often volumetric imaging such as CT or magnetic resonance imaging (MRI), and the patient. To maintain registration and synchronization of the viewpoint in the user's perspective, the position and orientation of the AR viewing camera with respect to the environment need to be tracked. 

For tracking and registration, two fundamental paradigms can be distinguished: \textit{outside-in} and \textit{inside-out} approaches. Outside-in (or extrinsic) tracking refers to strategies where external sensors (e.g., cameras) are stationed around the user and thus, observe the movement of the device from the outside. Such methods can be very accurate, but require many components and only work in a limited space. In inside-out (or intrinsic) tracking, the sensors are integrated within the AR device itself and thus, the device can self-locate within an unprepared environment. Although diverse types of sensors can be used for tracking, vision-based methods, relying on visible light, infrared (IR) cameras or depth sensors, have dominated the field for many years \cite{zhou2008trends}. For vision-based tracking, observable features need to be visible to the tracking cameras. Typically, these features can be divided into artificial features for \textit{marker-based} tracking, and natural features for \textit{marker-less} tracking.

Marker-based tracking relies on indicators of pre-defined pattern and size, whose location in relation to the real world is precisely known. These indicators can, for example, be fiducial markers visible by standard RGB cameras, or IR emitters (either active or passive), which are more robust to variable lighting conditions. Medical technology has appropriated this principle years ago: IR emitting markers are well-established in surgical navigation systems, where they are anchored in rigid tissue, such as the patients' bones, and on surgical instruments, while being tracked with stereo IR cameras. This approach allows a computation of the relative position of tools in relation to critical anatomy. 

Marker-less systems do not require artificial objects and, instead, rely on naturally observable features. Simultaneous localization and mapping (SLAM)~\cite{durrant2006simultaneous} and its variants are the most common markerless tracking techniques, which are capable of fusing information from various sensors (e.g., visible light, depth, GPS) to build a map of the environment and tracking the device within it. Virtual content can then be placed manually or with the aid of markers into the mapped world. Other marker-less tracking approaches involve models or templates of known, stationary real-world objects, which are then fitted to their real counterparts, either through 2D-3D (in case of visible light cameras) or 3D-3D (if 3D information of the scene is available) matching. Since 3D models of the patient's skin surface are typically available from medical imaging, such methods are well-suited for medical applications.

\subsection{The HoloLens}
The first generation HoloLens is wearable computer glass (often also referred to as "smartglass"), which delivers augmented reality experiences through a 3D optical see-through head-mounted display (OST-HMD). It was developed by Microsoft and rolled out in 2016. Contrary to other mixed or virtual reality headsets, the HoloLens was the first one to work fully untethered, requiring no wired connections to stationary infrastructure or prepared environments \cite{hl_hardware}.

The HoloLens features a set of built-in sensors, including an inertial measurement unit (IMU), four side-facing visible light cameras for capturing the environment, a time-of-flight (ToF) depth sensor, an ambient light sensor, four microphones and a front-facing, high definition photo/video camera. Only microphone and photo/video camera were accessible to developers in the beginning. In mid 2018, however, the so-called \textit{Research Mode} enabled access to ToF and environmental understanding cameras for research purposes~\cite{hl_rm}. Stereoscopic virtual content is displayed on two semi-transparent combiner lenses in front of the user's eyes for 3D vision, combined with the real environment. The equivalent of two 720p displays, one in front of each eye, allows a diagonal field of view (FOV) of 34 degrees, with a resolution of 47 pixels per degree~\cite{hl_fov}. Sound is delivered via built-in speakers. The HoloLens is equipped with an Intel Atom x5 32-bit central processing unit (CPU) with 1 GB of random access memory (RAM), and has 64 GB of storage. Its active battery life is specified at 2-3 hours.

A custom, dedicated hardware accelerator, the Holographic Processing Unit with 1 GB of additional RAM, enables efficient processing of the sensor data in parallel to processes running on the HoloLens' CPU. This custom chip facilitates a set of on-board capabilities to understand the users actions, as well as the environment around the device. A SLAM algorithm continuously constructs and refines a spatial map of the environment, and locates the device within it, resulting in on-board, marker-less inside-out tracking of the HoloLens~\cite{Klein2017}. Gaze tracking is supported via analyzing the user's head movement. Users can interact with virtual content via hand gestures or voice commands, both of which are automatically recognized. Additional input devices can be connected to the device via Bluetooth 4.1 LE, for example, the included clicker, a gamepad or an external keyboard. Connections can further be established wireless via Wi-Fi 802.11ac, or wired via Micro USB 2.0. 


\section{Methodology}
\subsection{Search strategy and selection process}
We conducted a systematic review of existing research about the HoloLens applied in medical scenarios. The review followed the Preferred Reporting Items on Systematic Reviews and Meta-Analysis (PRISMA) guidelines by Moher et al.~\cite{moher2009preferred}. A systematic literature search in the databases PubMed and Scopus was performed for the keyword \textit{[hololens]}, together with any of the terms \textit{[medicine]}, \textit{[surgery]} or \textit{[healthcare]} in March 2022. The publication period was restricted to the years 2016 to 2021. Duplicates were removed, then, an initial screening of titles and abstracts was performed. After the initial screening, full texts were retrieved and reviewed for eligibility. Criteria for inclusion in both phases of screening were: 1) studies with English full texts, 2) studies describing full original research by the authors, 3) studies which have been peer reviewed, and 4) studies describing the application of the HoloLens primarily for a human medical purpose. Consequently, exclusion criteria were: 1) studies without English full texts, 2) studies not describing full original research, such as reviews or book chapters 3) studies which have not been peer-reviewed, for example conference posters/abstracts or commentaries, 4) studies which do not use the HoloLens as main AR device, but only mention it, and 5) studies which are not primarily focused on a human medical purpose, but on other applications such as industry or gaming, and only mention medicine as a possible field of application. \\

\begin{figure}
\centering
\small{
\begin{tikzpicture}[
    node distance=1em and 0.5em,
    start chain=going below,
 mynode/.style = {
        draw, rectangle, align=center, text width=3.5cm,
        font=\small, inner sep=2ex, outer sep=0pt,
        on chain},
mylabel/.style = {
        draw, rectangle, align=center, rounded corners, 
        font=\small\bfseries, inner sep=2ex, outer sep=0pt,
        fill=cyan!30, minimum height=38mm,
        on chain},
every join/.style = arrow,
     arrow/.style = {very thick,-stealth}
                    ] 
\coordinate (tc);
\node (n1a) [mynode, left=of tc]    {Records identified through database searching \\
                                        (975)};
\node (n1b) [mynode,right=of n1a]    {Additional records identified through other sources \\
                                        (18)};
\node (n2)  [mynode, below=of n1a]   {Records screened \\
                                        (993)};
\node (n3)  [mynode,join]   {Records after duplicates removed \\
                                (949)};
\node (n4)  [mynode,join]   {Full-text articles accessed for eligibility \\
                                (231)};
\node (n5)  [mynode,join]   {Studies included \\
                                (\numpapers)};
\node (n3r) [mynode,right=of n3]    {Records excluded, based on exclusion criteria \\
                                        (718)};
\node (n4r) [mynode,right=of n4]    {Full-text articles excluded, based on exclusion criteria \\
                                        (60)};
\draw[arrow] (n1a.south) coordinate (a) -- (a |- n2.north);
\draw[very thick, -stealth] (n1b.south) coordinate (b) -- (b |- n2.east) -> (n2);
\draw[arrow] (n3) -- (n3r);
\draw[arrow] (n4) -- (n4r);
\end{tikzpicture}
}
\caption{Search strategy used in this systematic review. Adapted from the PRISMA flow diagram by Moher et al.~\cite{moher2009preferred}.}
\label{tikz:prisma_flowchart}
\end{figure}
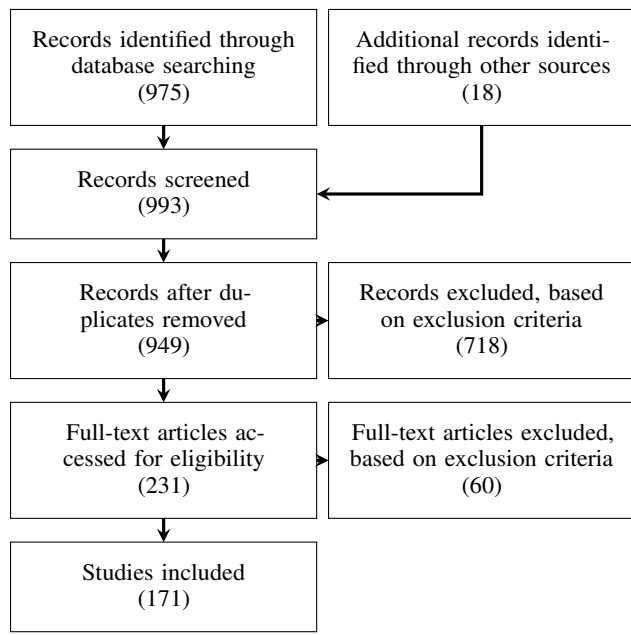

The systematic electronic search resulted in a total of 975 records. 18 additional records previously known to the authors were also considered. After removal of duplicates and screening of titles, abstracts and full texts according to our inclusion criteria, \numpapers~studies were selected for the final analysis (see \autoref{tikz:prisma_flowchart}). 

\subsection{Data extraction and taxonomy}
Each study was reviewed by one author. We extracted information about authors, year of publication and medical speciality from every publication. Medical specialities were determined as stated by the authors, by publication venue or targeted anatomy and grouped, were applicable, e.g., cranial and facial sub-specialities were combined as cranio-maxillofacial. Then, we extracted information about every publication according to our taxonomy, seen in~\autoref{fig:taxonomy}. In \autoref{sec:application}, we classified each study by the main intended user of the HoloLens: 1) clinical systems, whose main purpose is the support of physicians and healthcare professionals in the clinical routine, 2) Educational works, which aid medical and healthcare students in their schooling and training, and 3) applications focused on treatment and rehabilitation, which aim at supporting patients during different stages of therapy and disease management. Further, we divided each main category into sub-categories, based on application areas. From every publication, we also extracted information about applied registration and tracking methodologies, if any (see~\autoref{sec:reg}), where we first categorized studies based on their tracking paradigm (manual vs. inside-out vs. outside-in), and further distinguished between marker-based and marker-less methods. Data and visualization techniques are reviewed in~\autoref{sec:viz}, where we define categories based on data source (medical vs. non-medical), data type (2D, 3D and other), as well as acquisition time. Finally, we analyzed how medical AR applications using the HoloLens have been evaluated, grouping studies according to their evaluation scenarios, and identified commonly used qualitative and quantitative measures in~\autoref{sec:eval}.

\tikzset{
  basic/.style  = {text width=2cm, font=\sffamily, rectangle},
  root/.style   = {basic, draw, drop shadow, rounded corners=2pt, thin, align=center, fill=white},
  level-2/.style = {basic, draw, drop shadow, rounded corners=6pt, thin, align=center, fill=white, text width=3cm},
  level-3/.style = {basic, draw, thin, align=center, fill=white, text width=1.8cm},
  level-4/.style = {basic, thin, align=left, fill=white, text width=2.8cm, node distance=0.8cm, font=\small\sffamily}
}

\begin{figure*}[!ht]
\centering
\normalsize{
\begin{tikzpicture}[
  level 1/.style={sibling distance=12em, level distance=5em},
  edge from parent/.style={->,solid,black,thick,sloped,draw}, 
  edge from parent path={(\tikzparentnode.south) -- (\tikzchildnode.north)},
  >=latex, node distance=1.2cm, edge from parent fork down]

\node[root] {\textbf{HoloLens 1 in medicine}}
  child {node[level-2] (c1) {\textbf{Use case}}}
  child {node[level-2] (c2) {\textbf{Registration \& Tracking}}}
  child {node[level-2] (c3) {\textbf{Data \\ \& Visualization}}}
  child {node[level-2] (c4) {\textbf{Evaluation}}};

\node [below of = c1, xshift=5pt, level-3] (c11) {Intended users};
\begin{scope}[every node/.style={level-4}]
    \node [below of = c11, xshift=20pt] (c111) {Physicians};
    \node [below of = c111] (c112) {Students};
    \node [below of = c112] (c113) {Patients};
\end{scope}

\node [below of = c113, level-3, xshift=-20pt] (c12) {Applications};
\begin{scope}[every node/.style={level-4}]
    \node [below of = c12, xshift=20pt] (c121) {Data display};
    \node [below of = c121] (c122) {Image-guided interventions};
    \node [below of = c122] (c123) {Surgical navigation};
    \node [below of = c123] (c124) {Intervention training};
    \node [below of = c124] (c125) {Anatomy learning};
    \node [below of = c125, text width=4cm, xshift=17pt] (c126) {Patient education \& training};
    \node [below of = c126, text width=4cm] (c127) {Assistance \& monitoring};
    \node [below of = c127, text width=4cm] (c128) {Diagnosis \& treatment};
\end{scope}

\node [below of = c2, xshift=5pt, level-3] (c21) {Paradigm};
\begin{scope}[every node/.style={level-4}]
    \node [below of = c21, xshift=20pt] (c211) {Manual};
    \node [below of = c211] (c212) {Inside-out};
    \node [below of = c212] (c213) {Outside-in};
\end{scope}
\node [below of = c213, xshift=-20pt, level-3] (c22) {Method};
\begin{scope}[every node/.style={level-4}]
    \node [below of = c22, xshift=20pt] (c221) {Marker-based};
    \node [below of = c221] (c222) {Marker-less};
\end{scope}

\node [below of = c3, xshift=5pt, level-3] (c31) {Data source};
\begin{scope}[every node/.style={level-4}]
    \node [below of = c31, xshift=20pt] (c311) {Medical};
    \node [below of = c311] (c312) {Non-medical};
\end{scope}
\node [below of = c312, xshift=-20, level-3] (c32) {Data type};
\begin{scope}[every node/.style={level-4}]
    \node [below of = c32, xshift=20pt] (c321) {3D};
    \node [below of = c321] (c322) {2D};
    \node [below of = c322] (c323) {Other};
\end{scope}
\node [below of = c323,  xshift=-20, level-3] (c33) {Acquisition time};
\begin{scope}[every node/.style={level-4}]
    \node [below of = c33, xshift=20pt] (c331) {Pre-interventional};
    \node [below of = c331] (c332) {Intra-interventional};
\end{scope}

\node [below of = c4, xshift=5pt, level-3] (c41) {Scenario};
\begin{scope}[every node/.style={level-4}]
    \node [below of = c41, xshift=20pt] (c411) {Proof-of-concept};
    \node [below of = c411] (c412) {Laboratory};
    \node [below of = c412] (c413) {Relevant environment};
\end{scope}
\node [below of = c413, xshift=-20pt, level-3] (c42) {Measures};
\begin{scope}[every node/.style={level-4}]
    \node [below of = c42, xshift=20pt] (c421) {Quantitative};
    \node [below of = c421] (c422) {Qualitative};
\end{scope}

\foreach \value in {1,2}
    \draw[->] (c1.190) |- (c1\value.west);
    
\draw[-] (c11.205) |- (c111.west);
\draw[-] (c11.205) |- (c112.west);
\draw[-] (c11.205) |- (c113.west);

\draw[-] (c12.197) |- (c121.west);
\draw[-] (c12.197) |- (c122.west);
\draw[-] (c12.197) |- (c123.west);
\draw[-] (c12.197) |- (c124.west);
\draw[-] (c12.197) |- (c125.west);
\draw[-] (c12.197) |- (c126.west);
\draw[-] (c12.197) |- (c127.west);
\draw[-] (c12.197) |- (c128.west);

\foreach \value in {1,...,2}
  \draw[->] (c2.200) |- (c2\value.west);
  
\draw[-] (c21.197) |- (c211.west);
\draw[-] (c21.197) |- (c212.west);
\draw[-] (c21.197) |- (c213.west);

\draw[-] (c22.195) |- (c221.west);
\draw[-] (c22.195) |- (c222.west);

\foreach \value in {1,...,3}
  \draw[->] (c3.200) |- (c3\value.west);
  
\draw[-] (c31.205) |- (c311.west);
\draw[-] (c31.205) |- (c312.west);

\draw[-] (c32.197) |- (c321.west);
\draw[-] (c32.197) |- (c322.west);
\draw[-] (c32.197) |- (c323.west);

\draw[-] (c33.205) |- (c331.west);
\draw[-] (c33.205) |- (c332.west);
  
\foreach \value in {1,2}
  \draw[->] (c4.185) |- (c4\value.west);
  
\draw[-] (c41.195) |- (c411.west);
\draw[-] (c41.195) |- (c412.west);
\draw[-] (c41.195) |- (c413.west);

\draw[-] (c42.195) |- (c421.west);
\draw[-] (c42.195) |- (c422.west);

\end{tikzpicture}
}
\caption{Taxonomy employed in this review. Each publication is analyzed with regard to intended use case, registration and tracking principles, data sources and visualization, as well as evaluation and validation.}
\label{fig:taxonomy}
\end{figure*}
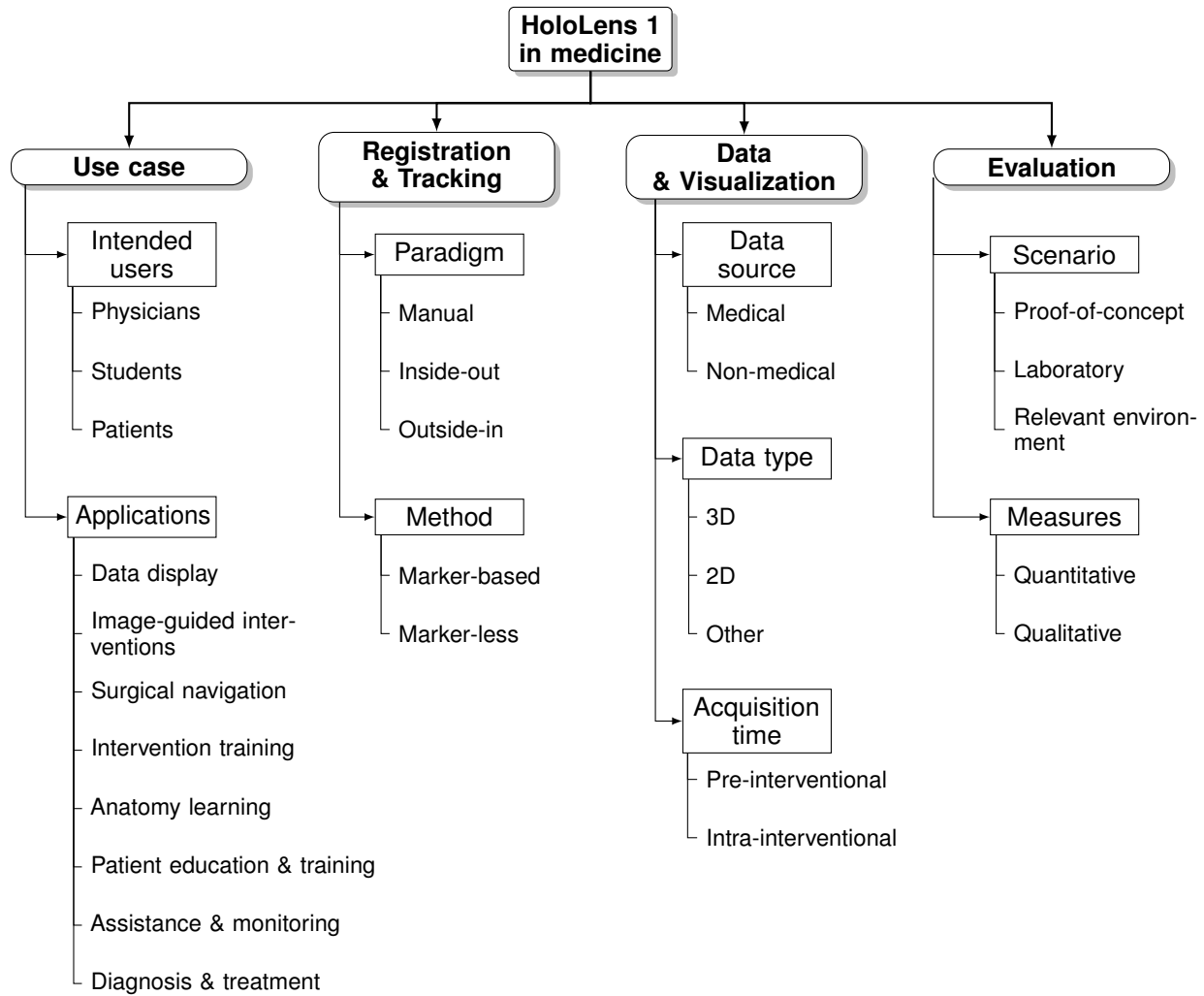

\subsection{Related reviews}

According to our exclusion criteria, review publications are not analyzed in this study. Still, we identified several related reviews during our literature search, which might be of interest for the reader.

Barsom et al.~\cite{barsom2016systematic} provide a systematic review about AR for medical training to the year of 2015, and found that, although promising results were achieved, full validation of training systems was lacking. Chen et al.~\cite{chen2017recent} analyze trends and challenges in medical AR found in over 1400 publications in the time period between 1995 and 2015. They identify powerful enabling technologies, human-computer-interaction and validation as major research challenges. Eckert et al.~\cite{eckert2019augmented} review medical AR applications described between the years of 2012 and 2017. In these years, a trend towards display technology research and medical treatment scenarios could be identified. Still, a lack of evidence in clinical studies was noted.

Several reviews about AR, specifically for \textit{surgical} applications, have been published. Vavra et al.~\cite{vavra2017recent} and Yoon et al.~\cite{yoon2018augmented} review articles published pre-HoloLens, between 2010 and 2016, as well as 1995 and 2017, respectively. In this time period, live streaming from endoscopy, followed by navigation and video recording, were the most popular applications. Rahman et al.~\cite{rahman2020head} focus specifically on HMD use in surgical scenarios up to the year of 2017. More recent reviews about surgical AR using OST-HMD come from Birlo et al.~\cite{birlo2022utility} and Doughty et al.~\cite{doughty2022augmenting} for the years between 2013 and 2020, and 2021 to March 2022, respectively. They clearly show that the Microsoft HoloLens was the major driving force in OST-HMD research for surgery in the past years. Even more specialized surgical reviews have been published for orthopedic surgery~\cite{jud2020applicability}, oral and cranio-maxillofacial surgery~\cite{badiali2020review,gsaxner2021archapter}, neurosurgery~\cite{meola2017augmented,guha2017augmented,lopez2019intraoperative}, laparoscopic surgery~\cite{bernhardt2017status} and robotic surgery~\cite{qian2019review}. 

In all these reviews, the lack of clinical validation is the most re-occurring aspect, something we also identify in this study. Other commonly mentioned challenges include technical limitations in regards to device tracking and rendering, and limited usability due to complicated workflows. The HoloLens, with its self-tracking capabilities, good support for the development of user interfaces and interactions and improved rendering capabilities, makes some of these challenges obsolete. Therefore, in this review, we focus exclusively on aspects and challenges coming with this new generation of OST-HMD devices, which still bear significance for more recent hardware, such as the HoloLens 2 or Magic Leap 2 (Magic Leap, Plantation, FL). Thus, we hope that it is interesting for not only looking back, but in particular also for pointing future researchers towards directions in which increased efforts are required.


\section{Publications per year}
\begin{figure}
    \centering
    \includegraphics[width=\columnwidth]{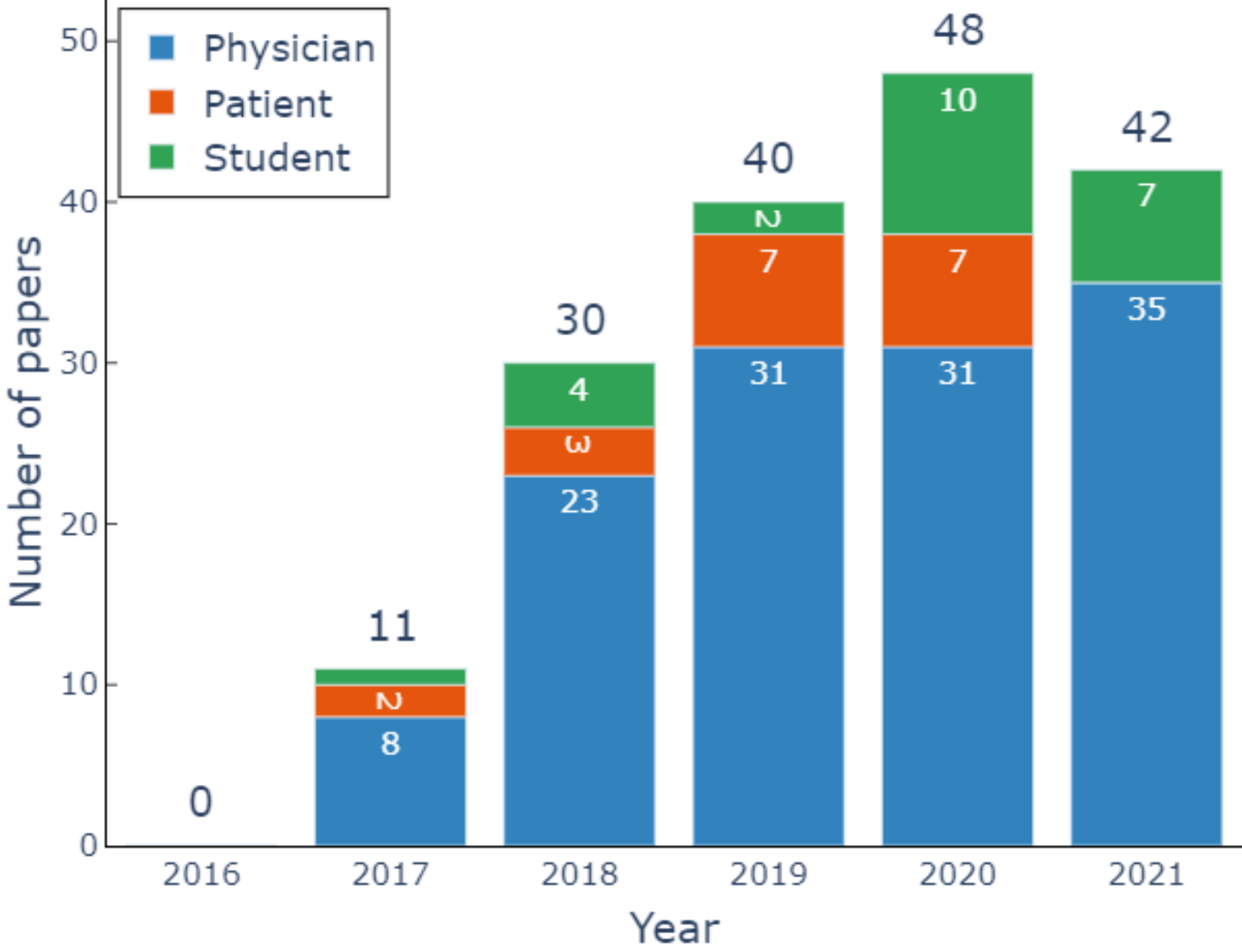}
    \caption{Number of papers published per year in each use case category between the years 2016 and 2021.}
    \label{fig:years_plot}
\end{figure}

Figure \ref{fig:years_plot} provides an overview of the number of papers published in each reviewed year, from 2016 to 2021. Although the HoloLens was available from March 2016 in North America and October 2016 worldwide, no publications reporting it's use in the medical domain were published in this year. After that, the number of publications in all categories shows a steady increase, with the highest number of research reported in 2020. In 2021, the number of papers decreases again -- likely caused by the release of the HoloLens 2, which lead many researchers to shift their attention towards the newer generation device. 

\section{Medical fields of applications}

\begin{figure}[ht]
    \centering
    \includegraphics[width=\columnwidth]{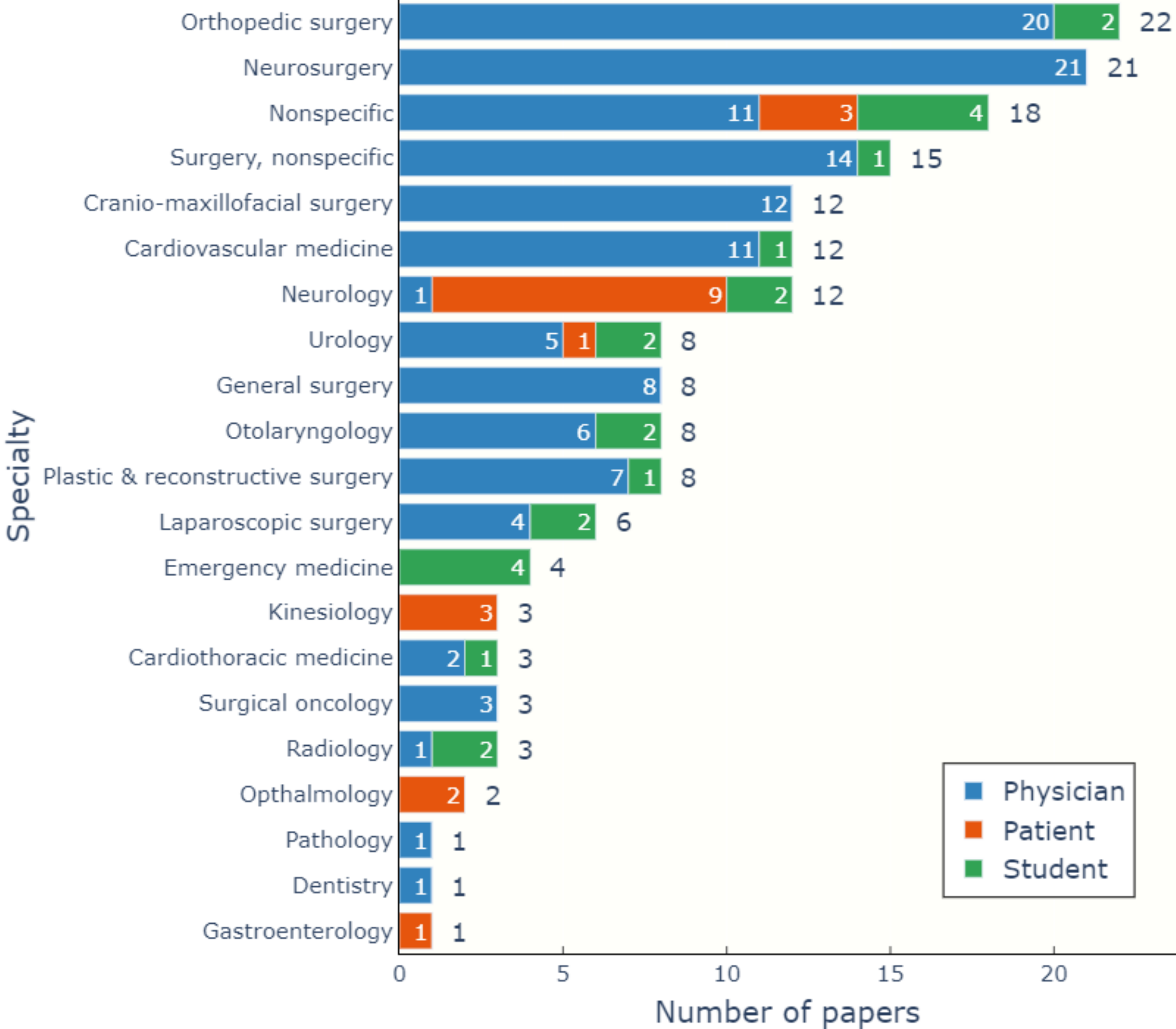}
    \caption{Frequency of papers in each of the 21 identified medical fields. "Nonspecific" refers to applications where authors did not indicate a specific area, which means they could be used in several disciplines.}
    \label{fig:specialty_plot}
\end{figure}

As shown in Figure~\ref{fig:specialty_plot}, the HoloLens saw applications in a large variety of medical areas, which we group into 21 fields. Surgical disciplines, in particular orthopedic surgery (18) and neurosurgery (14), were most frequently supported by AR applications, especially those targeted at physicians. Interestingly, in these most frequent disciplines, image-guided and navigated interventions are already particularly common, e.g., through surgical navigation systems or fluoroscopy. Hence, it can be assumed that, from the perspective of user acceptance and recognition, the translation of AR technology into clinical practice can be more successful in areas which already heavily rely on such technological assistance. However, relevant procedures have also highest demands in accuracy and safety, which makes the implementation of AR much more difficult from a technical standpoint. 15 publications do not indicate a specific medical field, and eleven target surgical procedures in general. These publications mostly introduce more general concepts not targeted at specific medical procedures -- thus, they could be used in more than one specialty. Patient-focused applications are rather situated in speciality areas, where patient cooperation and motivation has a large impact on treatment outcome, such as neurology and kinesiology. 

\section{Use cases}
\label{sec:application}

\begin{figure}
    \centering
    \includegraphics[width=\columnwidth]{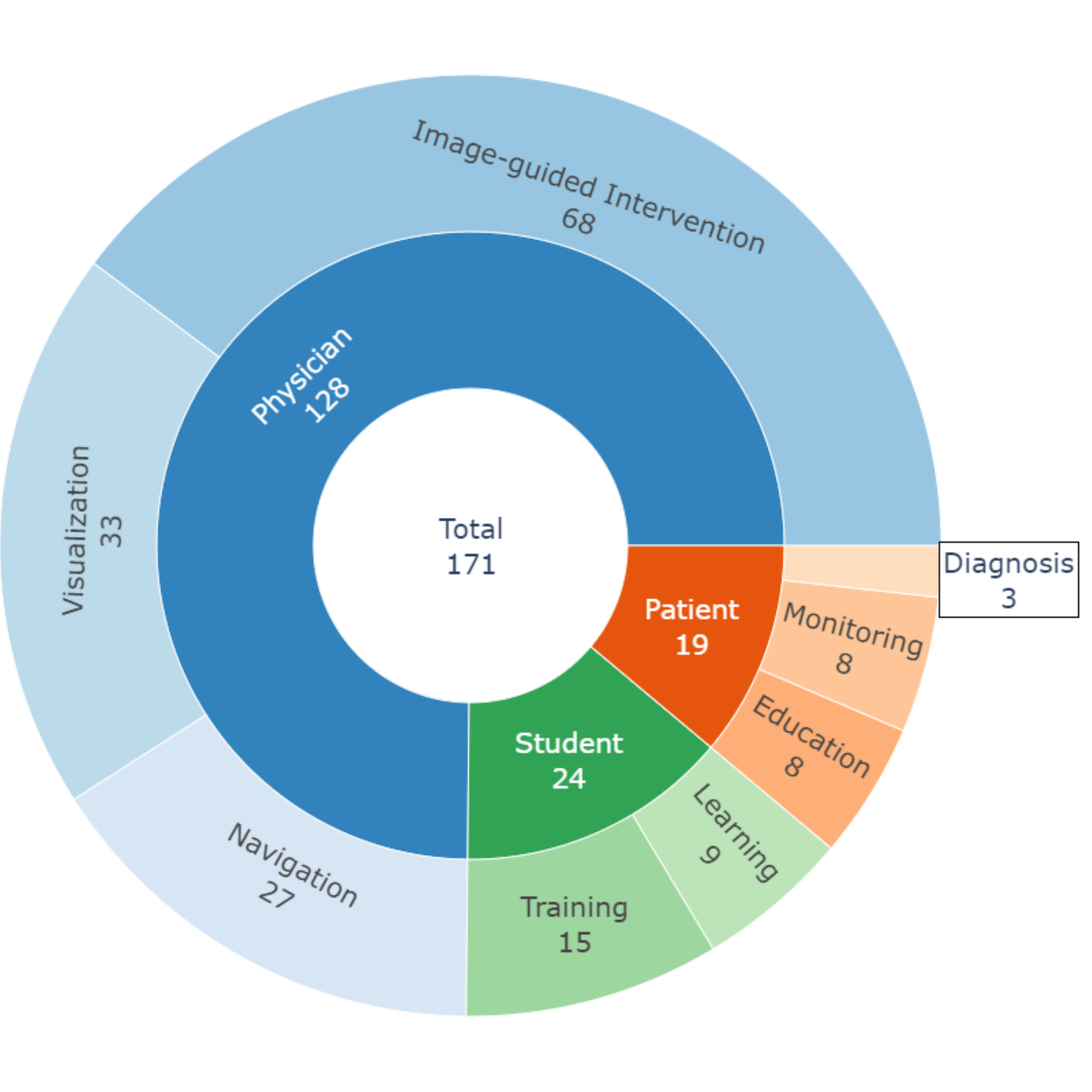}
    \caption{Overview of the number of papers identified in each of our three main categories (defined by targeted users) and sub-categories (defined by application area).}
    \label{fig:categories_plot}
\end{figure}

We first categorize publications by their intended users, and further by the supported application. An overview of the identified categories and number of associated publications is given in Figure~\ref{fig:categories_plot}. Physicians and healthcare professionals working within the clinical routine have been, by far, the most popular target audience of proposed HoloLens-based AR systems. 128 out of \numpapers~studies, almost 75\%, describe an application of the device for supporting healthcare professionals in tasks such as diagnosis, treatment planning and treatment execution. Medical students come second, with 24 works dedicated to anatomy learning or training of interventional procedures. Lastly, 19 studies targeted an application for patients, either for patient education, monitoring and guidance, or diagnosis. 

\subsection{Physician-centered applications of the HoloLens}
We group research within this category based on application, ranked by technological complexity: 1) Data visualization applications, where the HoloLens primarily serves as a display, are relatively simple to implement. 2) Image-guided interventions demand either a registration between virtual content and the patient or a way to display intra-operative imaging in real-time and are, consequently, more challenging. 3) Surgical navigation applications require tracking of medical tools in addition to the patient and the HoloLens, and have the highest demands in accuracy and reliability, which makes them the most complex. \autoref{tab:clinical} shows all studies targeted at physicians and other healthcare professionals, including their applications.

\begin{figure*}[ht]
  \centering
  \parbox{\textwidth}{
    \parbox{.3\textwidth}{%
      \subcaptionbox{}{\includegraphics[width=\hsize]{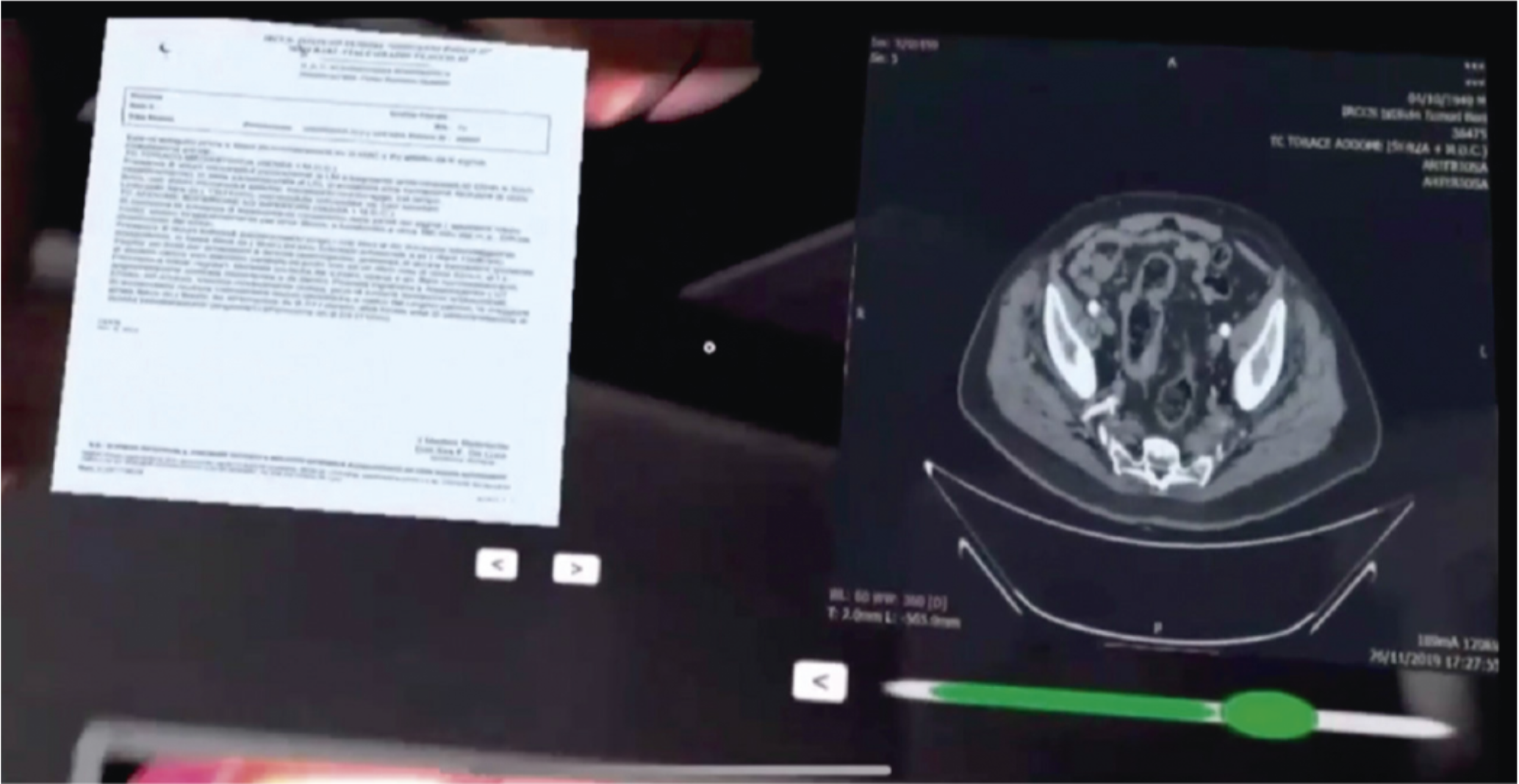}}
      \vskip0.5em
      \subcaptionbox{}{\includegraphics[width=\hsize, trim=0cm 1cm 0cm 1cm, clip]{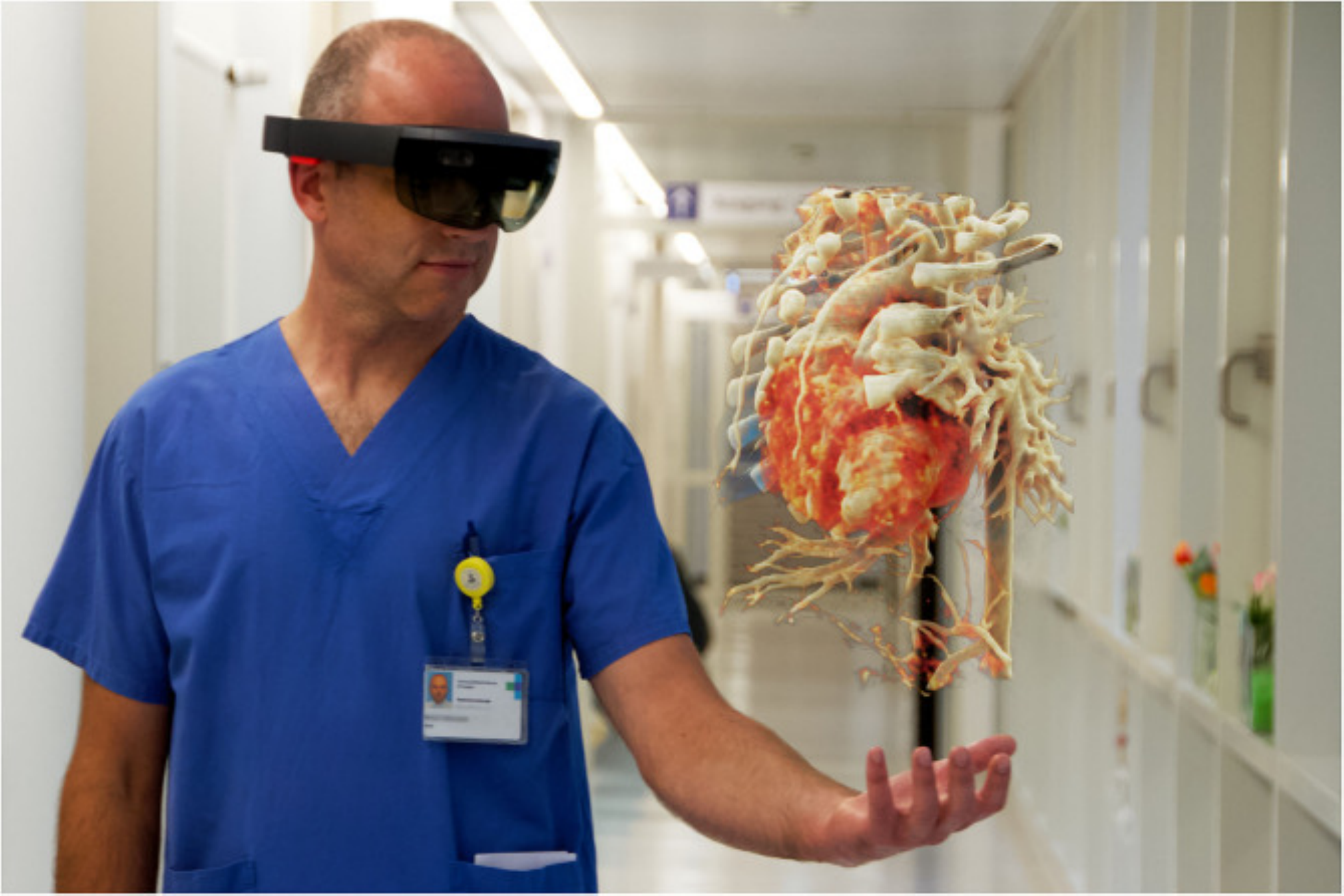}}
    }
    \hskip0.5em
    \parbox{.33\textwidth}{%
      \subcaptionbox{}{\includegraphics[width=\hsize, trim=1cm 0cm 1cm 0cm, clip]{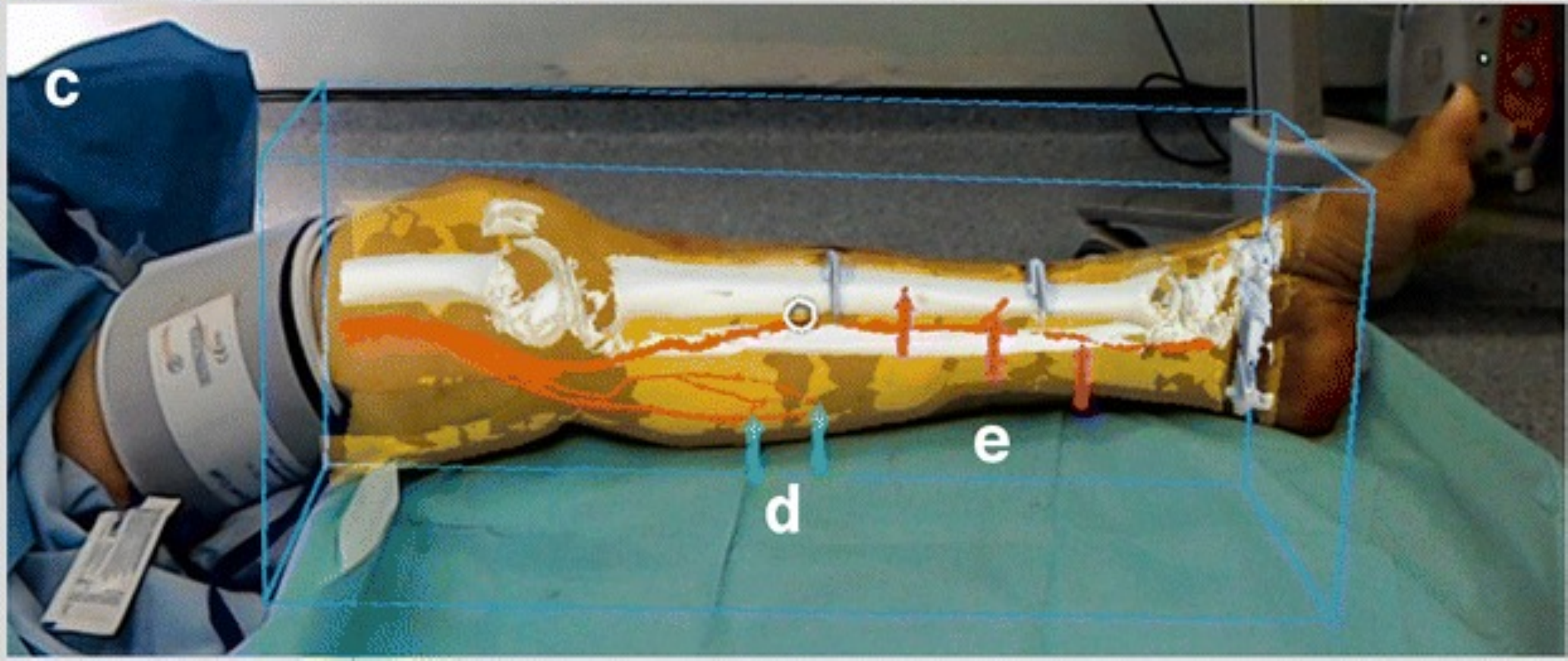}}
      \vskip0.5em
      \subcaptionbox{}{\includegraphics[width=\hsize, trim=1.9cm 0cm 1.5cm 0cm, clip]{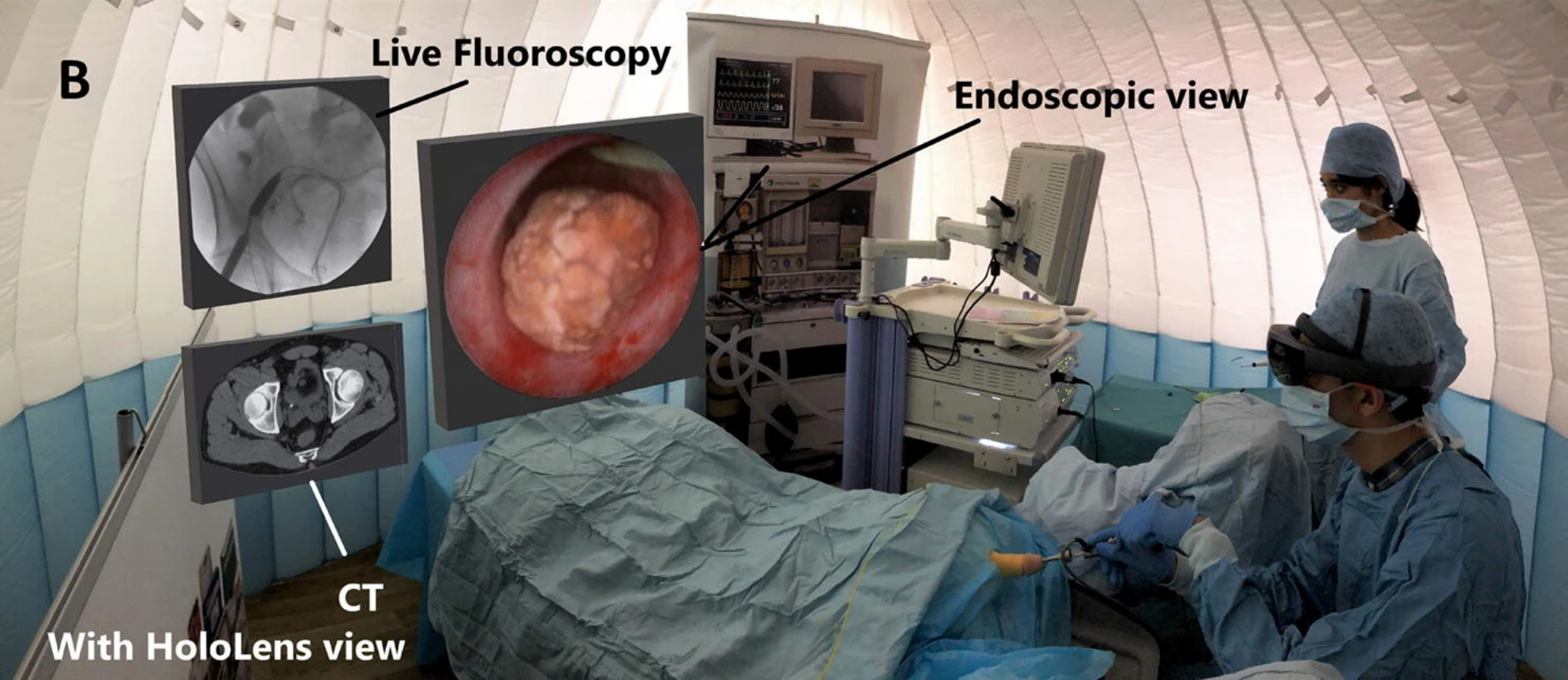}} 
    }
    \hskip0.5em
    \parbox{.34\textwidth}{%
      \subcaptionbox{}{\includegraphics[width=\hsize, trim=8cm 0cm 4.3cm 0cm, clip]{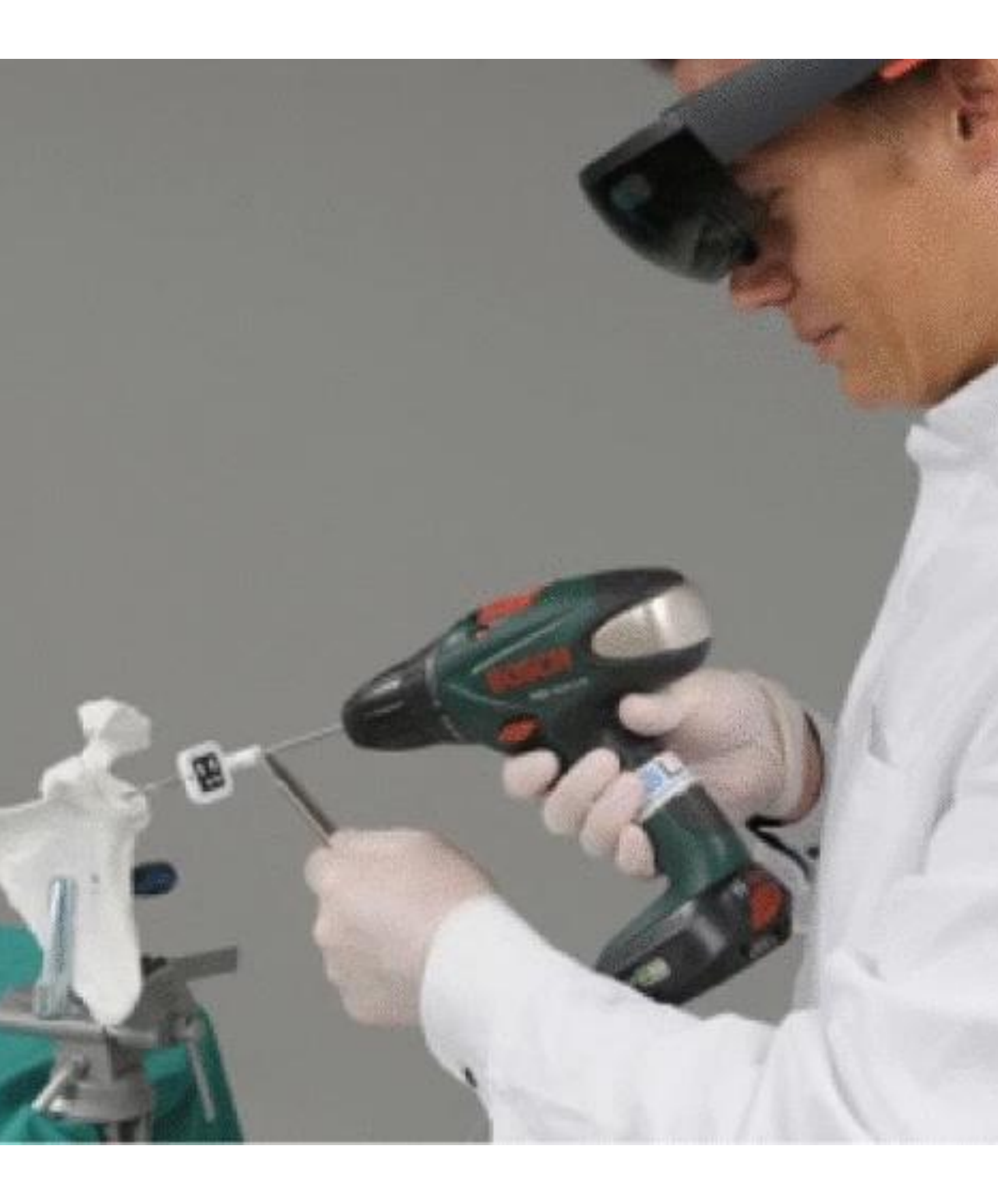}}
    }
  }
  \caption{Examples for physician-centered applications of the HoloLens. Left: Data display examples, showing immersive (a) 2D (Source: Galati et al.~\cite{galati2020experimental}, CC-BY) and (b) 3D patient data (Source: Gehrsitz et al.~\cite{gehrsitz2021cinematic}, CC-BY), without a reference to the physical space. Middle: Image-guided intervention via (c) X-ray vision (Source: Pratt et al.~\cite{pratt2018through}, CC-BY) and (d) live imaging (Source: Al-Janabi et al.~\cite{al2020effectiveness}, CC-BY). Right: (e) example for a surgical navigation application, were tools are tracked in relation to the target anatomy (Source: Kriechling et al.~\cite{kriechling2020augmented}, CC-BY)}
  \label{fig:physician}
\end{figure*}

\subsubsection{Data display}
In its simplest form, the HoloLens can be used as an immersive display for medical data, such as 2D/3D imaging or healthcare records (see~\autoref{fig:physician} (a) and (b)). Pure data display applications do not need to establish a correspondence between the physical space and the shown data -- content can simply be anchored to a fixed position according to the display itself, to be always visible for the wearer. Since the HoloLens self-locates within its environment, virtual objects can further be anchored to a stationary position within the real world without additional expenditure, to be naturally examined from different perspectives. This ability can have several advantages for clinicians. Access to medical data can be detached from stationary infrastructure and brought to treatment rooms, operating theaters and the bedside of the patient. For 3D data, such as volumetric medical imaging, stereoscopic visualization  through the HoloLens may lead to an improved perception of 3D relations. Furthermore, the possibility of touch-less interaction with data is ideal for scenarios where sterility is important. Finally, by synchronizing several headsets, visualizations may be more easily shared between users. These factors could make inspection of interaction with medical data during diagnosis, intervention planning and procedures more intuitive and less cumbersome. 
We identified 33 publications in this category. Most of them describe workflows for visualizing pre-interventionally acquired, 3D volumetric imaging data, such as CT, MRI or positron emission tomography (PET), but also healthcare records and other documents.

A smaller group of works explores telemedicine, where remote monitoring and assistance are important concepts. A remote expert can assist local staff in carrying out critical interventions, which is particularly useful in rural or disadvantaged areas, with limited funding and staff. The HoloLens features video conferencing capabilities, which enable the real-time transmission and visualization of the viewpoint of an interventionist to a remote expert/observers, and, vice versa, expert guidance via voice, video or annotations, without having to look away from the patient or using an external computer. Sirilak et al.~\cite{sirilak2018new} developed an e-consulting platform to connect specialized physicians with rural and remote hospitals. The feasibility of video and voice communication during intervention or surgery has further been explored by Mitsuno et al.~\cite{mitsuno2019telementoring} and Glick et al.~\cite{glick2020augmenting}. Proniewska et al.~\cite{proniewska2020three} developed a strategy for digitizing the operating room, allowing tele-monitoring from different perspectives with the HoloLens.

\subsubsection{Image-guided interventions}
The majority of papers reviewed in this study describe an application in image-guided intervention (IGI). AR for IGI is mainly motivated by the desire to obtain \textit{X-Ray vision} of a patient, which can incorporate medical imaging data intuitively into interventional workflows by aligning patient anatomy, imaging data and the physician's viewpoint. AR technology can superimpose pre- or intra-operative images and planning data directly with the patient, allowing the physician to see target structures through skin or obstructive anatomy, as seen in~\autoref{fig:physician} (c). It can, thus, either replace traditional image guidance, or provide guidance for interventions usually performed without.

Especially minimally-invasive interventions, which are performed without gaining direct access to the underlying anatomy, can benefit from X-ray visualization. Examples include skull base surgery ~\cite{mcjunkin2018development,kalavakonda2019augmented,creighton2020early}, arthroplasty~ \cite{agten2018augmented,wang2019hololens}, percutaneous orthopedic screw placement \cite{gibby2019head,liu2019percutaneous,buch2021development,dennler2021augmented}, ventricular drain insertion \cite{li2018wearable,rae2018neurosurgical,huang2019shared,schneider2021augmented} or ablations~\cite{ferraguti2020augmented,condino2021hybrid}. 

But X-ray visualization with the HoloLens has also been applied for procedures where the target anatomy is surgically exposed, such as tumor removal \cite{perkins2017mixed,incekara2018clinical,rose2019development,soulami2019mixed,huang2020augmented,saito2020intraoperative,ivan2021augmented,scherl2021augmented,scherl2021baugmented,gouveia2021breast}, vessel surgery~\cite{pratt2018through,katayama2020intraoperative,wesselius2021holographic}, or cranio-maxillofacial surgeries~\cite{koyachi2021accuracy,sugahara2021mixed,meng2021feasibility}. In these scenarios, the visualization of critical anatomical structures, which are not directly or clearly visible on the surgical site, such as blood vessels and nerves, or important planning information; for example, tumor resection margins or osteotomy lines, have the potential to make interventions safer. 

For a convincing X-Ray visualization, an accurate overlay of imaging data with the patient is a prerequisite. Image-to-patient-registration, relating virtual content with target anatomy, is the key component for such a system, but other factors, such as display calibration and stability of the HoloLens self-tracking also play an important role. While many of the aforementioned works rely on a manual alignment of virtual content with the patient, several publications within this category focus on addressing these technical challenges. Mostly, they do not focus on specific medical applications, but develop new concepts for system calibration~\cite{andress2018fly,hajek2018closing,fotouhi2019interactive} or  image-to-patient-registration~\cite{wu2018augmented,chien2019hololens,pepe2018pattern,sylos2019depth,gsaxner2019markerless}, which could be applied in various medical scenarios. Other works evaluate and compare selected technical aspects~\cite{frantz2018augmenting,mitsuno2019effective,van2019clinical,gu2021feasibility,perez2021effect}. We will discuss image-to-patient registration and calibration methods in more detail in section~\ref{sec:reg}

A third category of works does not target X-ray visualization -- instead, the HoloLens is used to enhance traditional image-guided interventions, such as laparoscopy, endoscopy, fluoroscopy or ultrasound. It has been shown that monitor placement during image-guided procedures plays an important role -- a misalignment of the visual-motor axis can increase fatigue and decrease orientation and hand-eye coordination of the operator, and, consequently, increase the risk of intervention-induced injuries \cite{el2006optimum}. By anchoring the virtual 2D "monitor" to a convenient physical location or the head gaze of the user, ergonomics and subjective workload may be improved. These applications require methods to deliver live medical data to the HoloLens in real-time. While most frameworks could support a variety of imaging sources, studies specifically evaluate intra-operative X-ray \cite{deib2018image,al2020effectiveness}, endoscopy \cite{al2020effectiveness}, ultrasound \cite{cartucho2020multimodal}, electro-anatomic mapping \cite{southworth2020performance} and MRI \cite{velazco2021modular}. An example is shown in~\autoref{fig:physician} (d).


\subsubsection{Surgical navigation}
Surgical navigation systems (SNS) have been shown to make procedures more accurate, less invasive and faster, resulting in improved outcomes for the patient~\cite{mezger2013navigation}. Compared to conventional image guidance using intra-operative X-ray or CT, SNS do not burden operators and patients with additional radiation exposure, and compared to ultrasound-based guidance, they are more accurate and work for every tissue type. Conventional SNS rely on visualizing navigation information on separate monitors, which leads to a \textit{switching focus problem} for surgeons -- they have to divide their attention between the surgical site and the navigation information. Such a division leads to issues of increased workload, disorientation and deteriorated hand-eye coordination~\cite{hansen2013auditory}, which AR could alleviate by fusing navigation information with the operating site. While IGI systems, as described above, can already provide a basic guidance based on images, precise surgical navigation requires real-time tracking of medical instruments and tools in relation to the patient anatomy, in addition to image-to-patient registration. In AR, navigation information can then be displayed \textit{in situ}, fused with the target anatomy, as shown in~\autoref{fig:physician} (e). 

Surgical navigation with the HoloLens has been explored as an alternative to commercial SNS in 27 publications. Mostly, AR navigation was studied in procedures where conventional SNS are already gold standard, such as neurosurgery~\cite{carbone2018proof,kunz2020infrared,van2021augmented,van2021effect}, orthopedic (in particular, spinal) surgery~\cite{el2018augmented,de2019hand,liebmann2019pedicle,kriechling2020augmented,muller2020augmented,spirig2021augmented}, general surgery~\cite{meulstee2019toward} or cranio-maxillofacial surgery~\cite{gao2019feasibility,sun2020fast,glas2021augmented}. AR SN can also provide an X-ray free alternative to interventions typically guided by intra-operative imaging, such as endovascular procedures~\cite{kuhlemann2017towards,garcia2018navigation,liu2019augmented} or tissue ablations~\cite{kuzhagaliyev2018augmented,li2019mixed}, or can be integrated into robotic surgery~\cite{qian2017comparison,qian2018arssist,qian2020flexivision}. 

These procedures have highest demands in accuracy and reliability of registration and tracking, with a high reference precision in a millimeter or sub-millimeter range. With the HoloLens hardware, it is difficult to meet these requirements. However, the HoloLens has a much slimmer form factor than conventional image guidance systems, which allows navigation for less critical procedures, usually performed without. Examples for such procedures include brain stimulation treatment~\cite{leuze2018mixed} and US examinations~\cite{ruger2020ultrasound,nguyen2022holous}.



Instrument tracking methods with the HoloLens will be reviewed in more detail in section~\ref{sec:reg}.

\begin{table*}[ht!]
  \centering
  \caption{Studies reporting an application of the HoloLens for physicians and other health care professionals.}
    \begin{tabularx}{\textwidth}{p{0.2\textwidth} p{0.15\textwidth} p{0.59\textwidth}}
    \toprule
    \multicolumn{1}{S}{Application} & Focus & \multicolumn{1}{X}{Studies} \\
    \midrule
        \multirow{2}{*}{Data display (33)} & \makecell[l]{ Medical Data \\ Visualization (29)}   &  \citet{mojica2017holographic,qian2017comparison,sauer2017mixed,bucioli2017holographic,frohlich2018holographic,jang2018three,affolter2019applying,brun2019mixed,checcucci20193d,kubben2019feasibility,moosburner2019real,soulami2019mixed,talaat2019three,witowski2019augmented,allison2020breast3d,bulliard2020preliminary,fitski2020mri,galati2020experimental,kumar2020use,pelanis2020use,perkins2020patient,cofano2021augmented,dennler2021baugmented,gehrsitz2021cinematic,iqbal2021augmented,morales2021holographic,saito2022intraoperative,velazco2021evaluation,wake2021workflow}\\
        \cmidrule{2-3}
        & \makecell[l]{Tele-medicine (5)}  &  \citet{sirilak2018new,mitsuno2019telementoring,glick2020augmenting,proniewska2020three,cofano2021augmented}\\
    \midrule
        \multirow{3}{*}{\makecell[l]{Image-guided \\ interventions (68)}} 
        
        & \makecell[l]{X-ray vision: \\ clinical focus (38)}    &  \citet{perkins2017mixed,agten2018augmented,hanna2018augmented,incekara2018clinical,li2018wearable,mcjunkin2018development,pratt2018through,rae2018neurosurgical,amini2019augmented,gibby2019head,huang2019shared,kalavakonda2019augmented,liu2019percutaneous,lohou2019preliminary,rose2019development,wang2019hololens,creighton2020early,ferraguti2020augmented,huang2020augmented,katayama2020intraoperative,nuri2020augmented,saito2020intraoperative,tian2020validation,buch2021development,condino2021hybrid,dennler2021baugmented,gouveia2021breast,iizuka2021potential,ivan2021augmented,koyachi2021accuracy,long2021comparison,meng2021feasibility,qi2021holographic,scherl2021augmented,scherl2021baugmented,schneider2021augmented,sugahara2021mixed,wesselius2021holographic}\\
        \cmidrule{2-3}
        & \makecell[l]{X-ray vision: \\ technical focus (24)}   &  \citet{xie2017holographic,andress2018fly,frantz2018augmenting,hajek2018closing,moreta2018augmented,pepe2018pattern,wu2018augmented,chien2019hololens,fotouhi2019interactive,gsaxner2019markerless,mitsuno2019effective,pepe2019marker,sylos2019depth,van2019clinical,fischer2020evaluation,jiang2020hololens,luzon2020value,nguyen2020augmented,nguyen2020baugmented,zuo2020novel,castelan2021augmented,gsaxner2021augmented,gu2021feasibility,perez2021effect}\\
        \cmidrule{2-3}
          & \makecell[l]{Live imaging (6)}    &  \citet{cui2017augmented,deib2018image,al2020effectiveness,cartucho2020multimodal,southworth2020performance,velazco2021modular}\\
    \midrule
    \multirow{2}{*}{Surgical navigation (27)} & \makecell[l]{}  &  \citet{kuhlemann2017towards,carbone2018proof,el2018augmented,garcia2018navigation,kuzhagaliyev2018augmented,leuze2018mixed,qian2018arssist,de2019hand,gao2019feasibility,li2019mixed,liebmann2019pedicle,liu2019augmented,meulstee2019toward,qian2019aramis,kriechling2020augmented,kunz2020infrared,muller2020augmented,qian2020flexivision,ruger2020ultrasound,sun2020fast,glas2021augmented,li2021augmented,liu2021wearable,nguyen2022holous,spirig2021augmented,van2021augmented,van2021effect}\\
    \bottomrule
    \end{tabularx}%
  \label{tab:clinical}%
\end{table*}%

\subsection{Applications of the HoloLens for medical students}
While the HoloLens 1 was originally not intended as a device for IGI or SN, its use as a tool for medical education was actively promoted. The CAE VimedixAR (CAE Healthcare, Montreal, Canada) app, an AR ultrasound training simulator, was amongst the first commercial applications available for the HoloLens 1, and tools for studying anatomy, such as HoloHuman by 3DMedical (Elsevier, Amsterdam, Netherlands) quickly followed. Probably due to the availability of commercial solutions, research in the area of medical education and training with the HoloLens is not as numerous as one might expect. We identified 24 publications in the area of HoloLens-based medical and healthcare student support, which we further categorize into: 1) Interventional and surgical training and 2) Anatomy learning. An overview is given in \autoref{tab:education}.

\begin{figure*}[ht]
  \centering
  \parbox{\textwidth}{
    \parbox{.54\textwidth}{%
      \subcaptionbox{}{\includegraphics[width=\hsize, trim=0cm 0cm 0cm 0cm, clip]{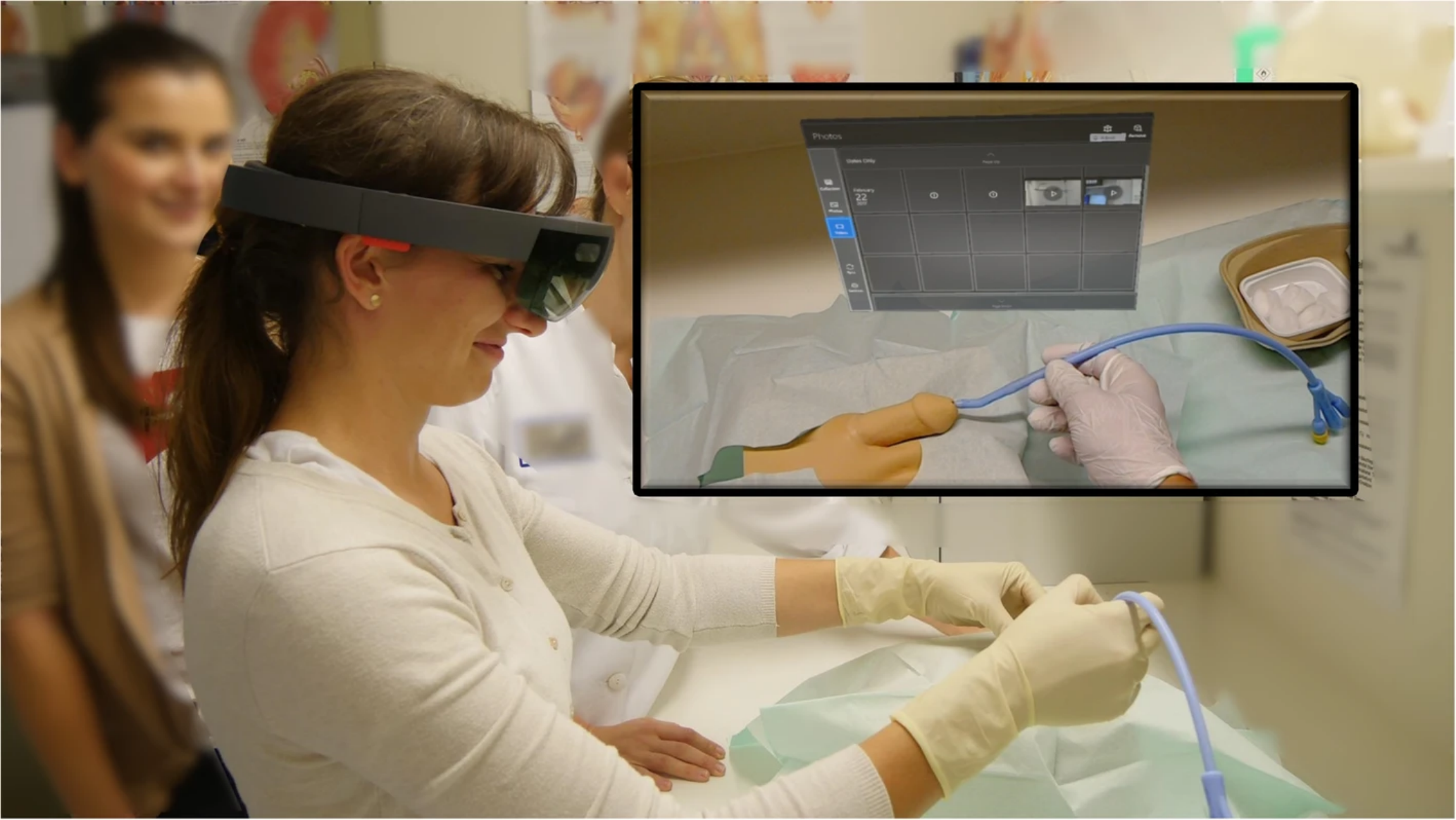}}
    }
    \hskip0.5em
    \parbox{.45\textwidth}{%
      \subcaptionbox{}{\includegraphics[width=\hsize, trim=0cm 0cm 0cm 0cm, clip]{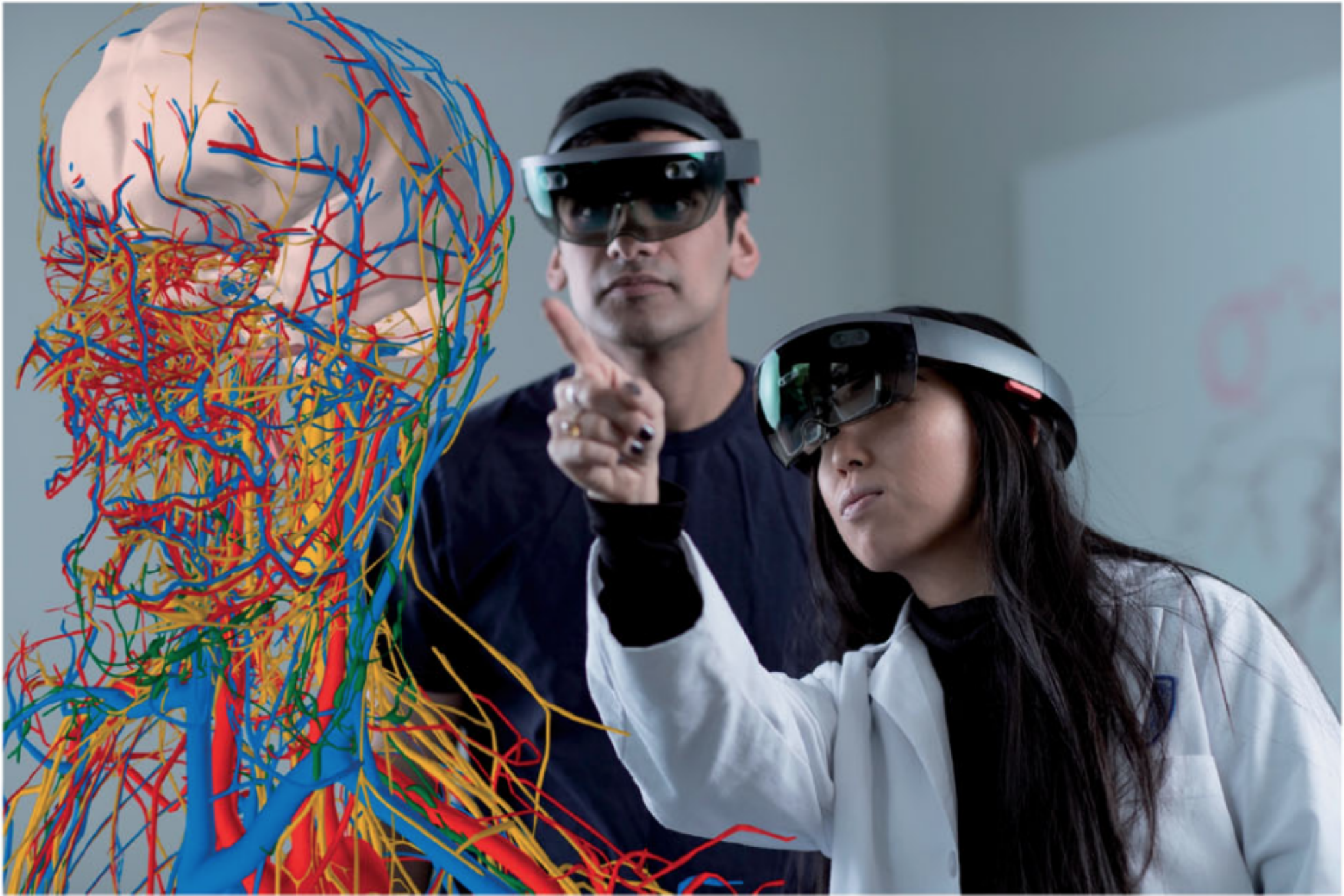}} 
    }
  }
  \caption{Examples of HoloLens applications for medical students. (a): A hybrid simulator for training catheter insertion with AR guidance (Source: Schoeb et al.~\cite{schoeb2020mixed}, CC-BY). (b) Studying anatomy with the HoloLens (Source: Ruthberg et al.~\cite{ruthberg2020mixed}).}
  \label{fig:student}
\end{figure*}

\subsubsection{Interventional and surgical training}
Simulation-based skill training has made its way into standard medical education, replacing or enhancing traditional teaching and training methods~\cite{gaba2004future}. Aside from traditional simulators based on physical manikins, mixed reality technology has gained considerable popularity in this domain, either by enabling fully virtual environments, or by enhancing manikin-based training through virtual guidance and feedback~\cite{so2019simulation}. 15 reviewed studies fall into this category.

The HoloLens 1 has been integrated into hybrid simulators, where it can be used to display additional guidance or even direct feedback to the user. Examples include the training of orthopedic surgery~\cite{cecil2018design,condino2018build}, emergency medicine interventions~\cite{kobayashi2018exploratory,balian2019feasibility,hong2020exploring,putnam2021virtual}, laparoscopic or US examinations~\cite{mahmood2018augmented,rewkowski2020small,heinrich2021holopointer} or urological procedures~\cite{muangpoon2020augmented,schoeb2020mixed}. Another possibility is to build fully simulated, virtual training scenarios~\cite{cecil2018design,brunzini2021mixed} or to include remote experts into the training sessions~\cite{wang2017augmented}.

\subsubsection{Anatomy learning}
A meta-survey by Yammine et al.~\cite{yammine2015meta} has shown that 3D visualization techniques are preferable over traditional methods for learning and teaching anatomy, both in terms of factual and spatial knowledge. Contrary to such visualizations on conventional monitors or in virtual reality (VR), AR could not only provide 3D visuals, but also annotate real, physical models or cadavers with digital information.

Studies evaluating the use of the HoloLens to teach various gross anatomy via 3D visualizations of anatomical models have been reviewed~\cite{stojanovska2019mixed,maniam2020exploration,antoniou2020biosensor,gnanasegaram2020evaluating,ruthberg2020mixed,moro2021hololens}. Robinson et al.~\cite{robinson2020evaluating} further tested the HoloLens as a learning platform for studying microscopic anatomy. 

\begin{table*}[ht!]
  \centering
  \caption{Studies reporting an application of the HoloLens for medical students and residents in an educational context.}
    \begin{tabularx}{\textwidth}{>{\hsize=0.4\hsize}X >{\hsize=1.6\hsize}X}
    \toprule
    Main application & \multicolumn{1}{l}{Studies} \\
    \midrule
    \makecell[l]{Interventional and \\ surgical training (15)}     &  \citet{wang2017augmented,cecil2018design,condino2018build,kobayashi2018exploratory,mahmood2018augmented,balian2019feasibility,hong2020exploring,lu2020integrating,muangpoon2020augmented,rewkowski2020small,schoeb2020mixed,brunzini2021mixed,heinrich2021holopointer,putnam2021virtual,suzuki2021learning} \\
    \midrule
    \makecell[l]{Anatomy learning (9)}     &  \citet{stojanovska2019mixed,antoniou2020biosensor,gnanasegaram2020evaluating,maniam2020exploration,robinson2020evaluating,ruthberg2020mixed,bogomolova2021development,kumar2021novel}\\
    \bottomrule
    \end{tabularx}%
  \label{tab:education}%
\end{table*}%

\subsection{Patient-focused applications of the HoloLens}
19 publications describe HoloLens-based systems for assisting patients during rehabilitation and treatment. Designing AR applications for patients is challenging due to age demographics, varying affinity to novel technologies and general anxiety when it comes to medical treatments. The novelty of AR technology also provides opportunities, since it can make otherwise repetitive or dull activities significantly more engaging.  We deduce three main application areas in this domain: a) patient training and education, b) assistance and monitoring and c) assessment and diagnosis. An overview over all studies, grouped by their application, is given in~\autoref{tab:patient}.

\begin{figure*}[ht]
  \centering
  \parbox{\textwidth}{
    \parbox{.32\textwidth}{%
      \subcaptionbox{}{\includegraphics[width=\hsize, trim=2cm 0cm 2cm 0cm, clip]{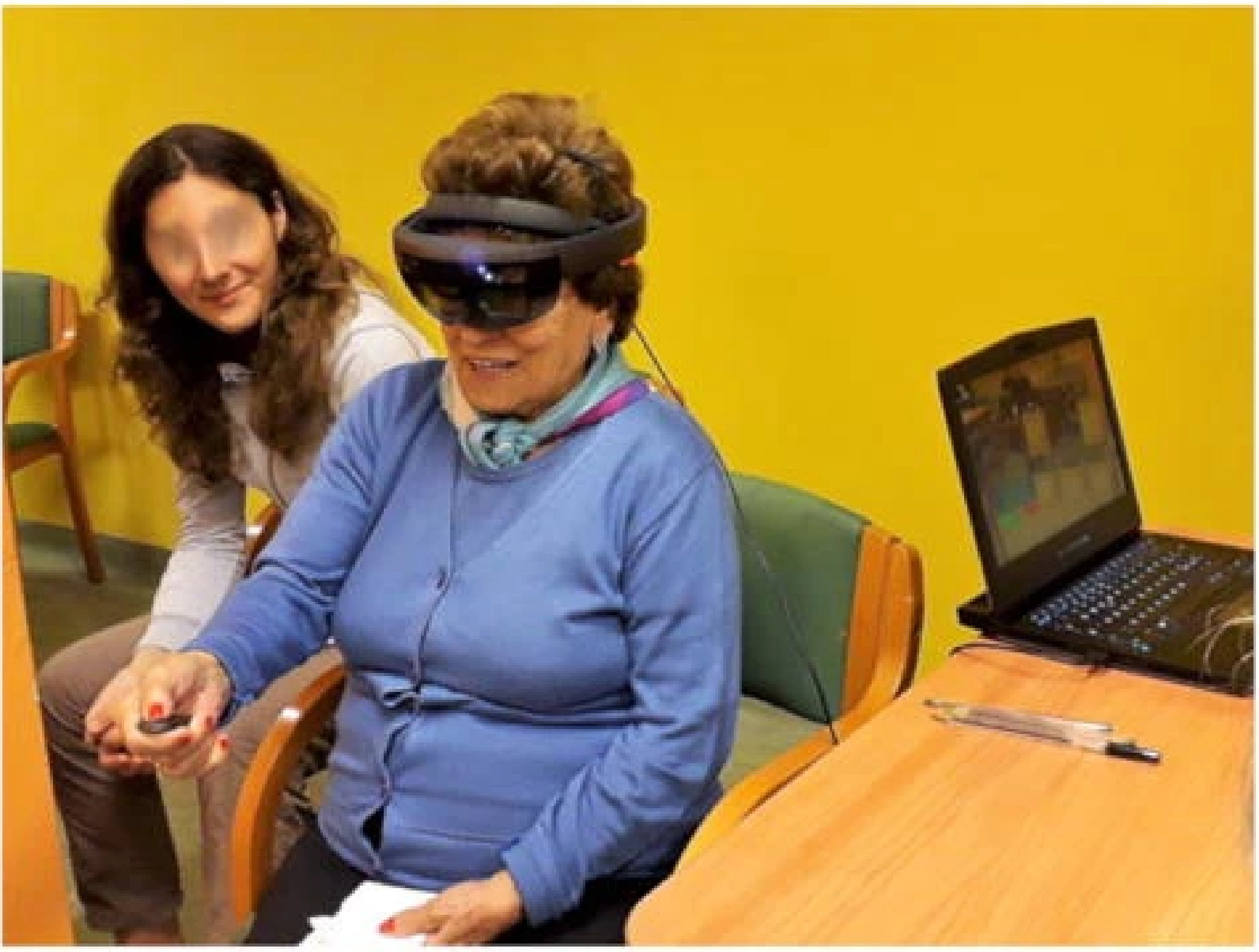}}
    }
    \hskip0.5em
    \parbox{.3\textwidth}{%
      \subcaptionbox{}{\includegraphics[width=\hsize, trim=0cm 1cm 0cm 1cm, clip]{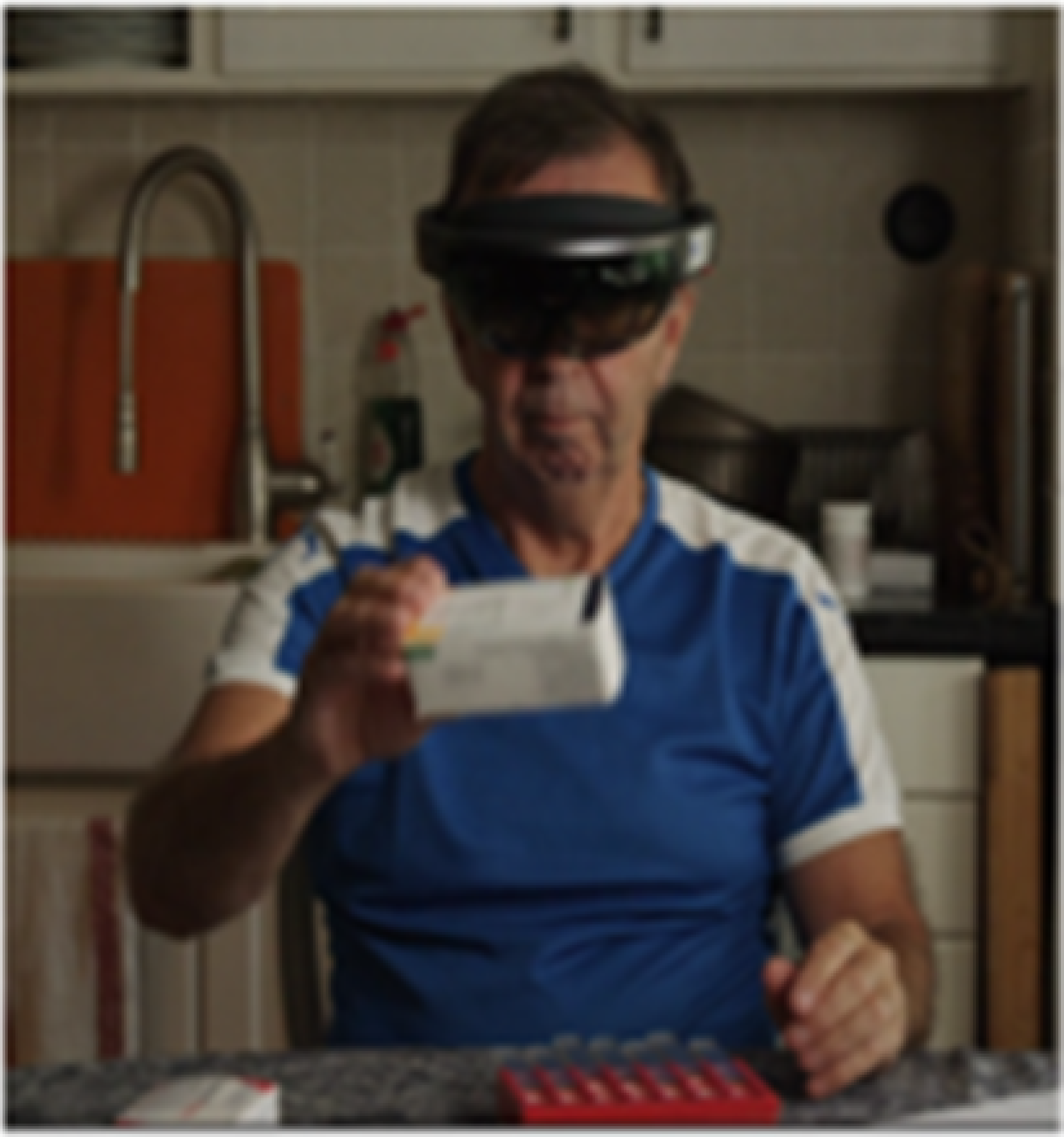}} 
    }
    \hskip0.5em
    \parbox{.35\textwidth}{%
      \subcaptionbox{}{\includegraphics[width=\hsize, trim=0cm 0cm 0cm 0cm, clip]{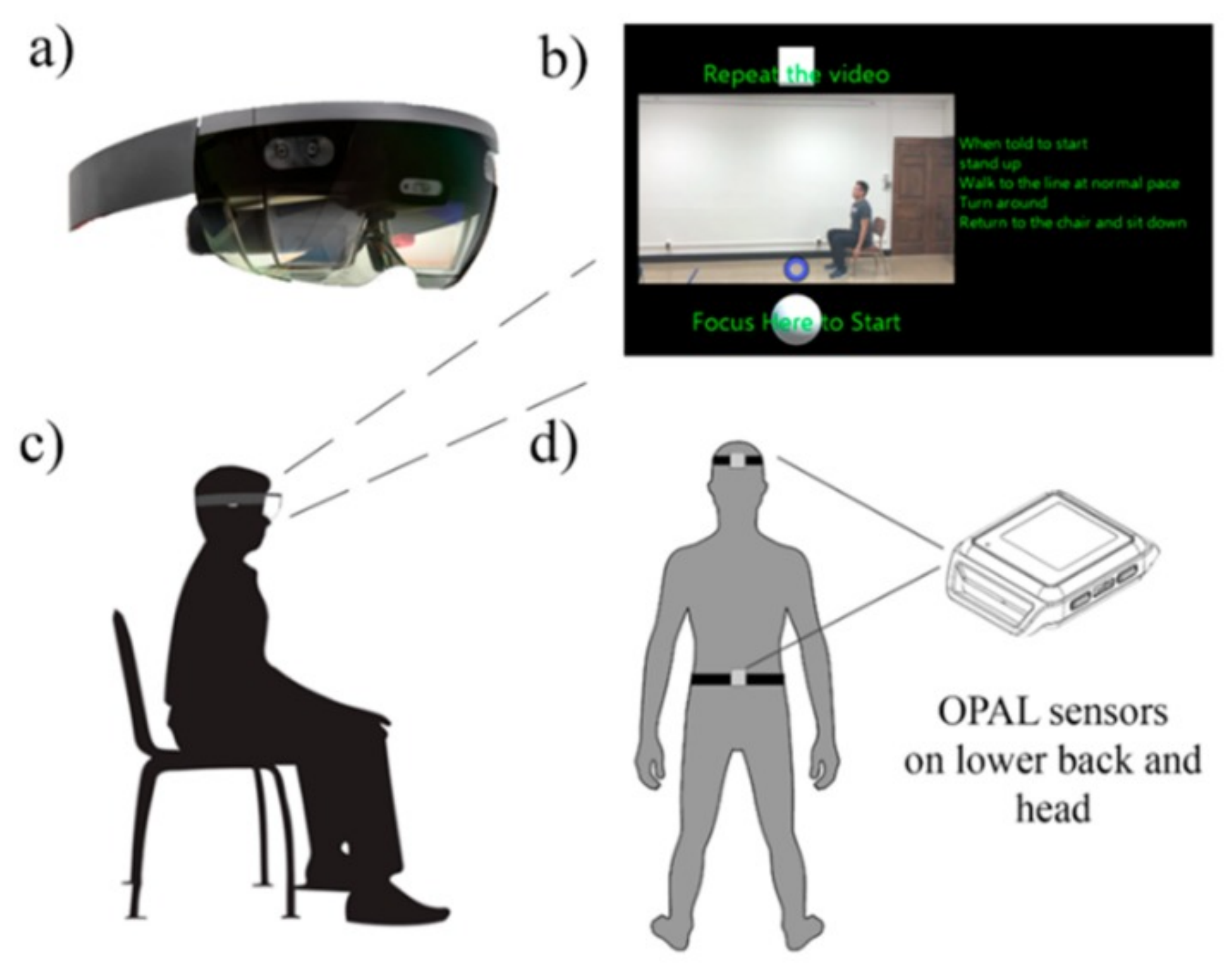}} 
    }
  }
  \caption{Example applications for the patient. (a): The HoloLens is used during a training session for Alzheimer's patients (Source: Aruanno et al.~\cite{aruanno2019memholo}). (b) The HoloLens as an assistant and monitoring tool for medication adherence (Source: Blusi et al.~\cite{blusi2019feasibility}, CC-BY-NC). (c) A HoloLens-based system for functional mobility assessment (Source: Sun et al.~\cite{sun2019validity}, CC-BY).}
  \label{fig:patient}
\end{figure*}

\subsubsection{Patient training and education}
It has been shown that immersive experiences can improve patient engagement and satisfaction during training tasks in rehabilitation~\cite{tieri2018virtual} and pre-interventional patient education~\cite{pandrangi2019application}. Therefore, AR environments have the advantage of being potentially more intriguing for patients than conventional methods. At the same time, AR scenarios are safe and easy to control. 

A series of studies has investigated the usage of the HoloLens to create virtual training environments for people with cognitive disorders, such as Alzheimer's disease~\cite{aruanno2017hololens,garzotto2018hololearn,aruanno2019memholo}. Another training task, which has benefited from AR support through the HoloLens, is the control of functional prostheses~\cite{sharma2018mixed,palermo2019augmented}. In the context of patient education, the HoloLens has been used to provide a more comprehensible and imaginable explanation to patients before surgery~\cite{wake2019patient,house2020use,rositi2021presentation}.

\subsubsection{Assistance and monitoring}
AR, with its ability to enhance the reality around the users in real-time, without insulating them, could be ideal for compensating various impairments and overcoming difficulties during the daily lives of patients. Mobile health (mHealth) applications support such procedures through mobile devices, such as smartphones, smartwatches, or, in this case, the HoloLens, and are, consequently, fitting for scenarios outside of a clinical environment, e.g., in the homes of patients.

The HoloLens has been explored for aiding patients with vision impairments in navigating their surroundings~\cite{yamashita2017pedestrian,angelopoulos2019enhanced}. Other applications include assisting patients with cognitive disorders in everyday activities~\cite{rohrbach2019augmented,janssen2020effects}, helping outpatients to adhere to their care plans~\cite{ingeson2018microsoft,blusi2019feasibility,boyd2020augmented} and text editing for people with motor disabilities~\cite{guerrero2020holonote}.

As mHealth applications are becoming more and more pervasive in our everyday lives, integrating them into augmented environments is a logical step, and the above-mentioned studies suggest promising applications of head-worn AR devices in mHealth. However, it should be noted that the HoloLens is not yet suitable for operation during everyday activities, as it is relatively expensive, and its short battery live and bulky form factor make it unfit for being worn and used for an extended period of time.

\subsubsection{Assessment and diagnosis}
The variety of built-in sensors, along with its self-tracking capabilities, unfold the possibility to utilize the HoloLens as a measurement device during patient assessments and diagnosis. At the same time, instructions and demonstrations, guiding patients through these tests, can be displayed immersively and interactively. 

Sun et al.~\cite{sun2019validity} used the HoloLens for leading and tracking patient performance during functional mobility tests, by evaluating the inertial measurement unit (IMU) data recorded by the device. Geerse~\cite{geerse2020quantifying} and Koop~\cite{koop2020hololens} utilize motion data collected by the HoloLens to assess gait parameters (e.g., walking speed, step length, cadence) in patients with movement disorders, in particular Parkinson's disease.

HoloLens-supported assessment and diagnosis is presumably closest to real clinical applicability in the domain of patient-oriented applications, as all reviewed studies have shown reliability of the measurements derived from the HoloLens sensors. At the same time, the ability to simultaneously monitor clinical parameters, while providing instructions to the patient with a single device has obvious benefits in terms of ergonomics and economics. Furthermore, using the HoloLens during the confined timespan of such screenings is feasible without undue discomfort for the patient. 

\begin{table*}[ht!]
  \centering
  \caption{Studies reporting a patient-focused application of the HoloLens.}
    \begin{tabularx}{\textwidth}{>{\hsize=0.4\hsize}X >{\hsize=1.6\hsize}X}
    \toprule
    Main application & Studies \\
    \midrule
    \makecell[l]{Patient training \\ and education (8)} &  \citet{aruanno2017hololens,garzotto2018hololearn,sharma2018mixed,aruanno2019memholo,palermo2019augmented,wake2019patient,house2020use,rositi2021presentation} \\
    \midrule
    \makecell[l]{Assistance \\ and monitoring (8)} &  \citet{yamashita2017pedestrian,ingeson2018microsoft,angelopoulos2019enhanced,blusi2019feasibility,rohrbach2019augmented,boyd2020augmented,guerrero2020holonote,janssen2020effects}\\
    \midrule
    \makecell[l]{Assessment \\ and diagnosis (3)}  &  \citet{sun2019validity,geerse2020quantifying,koop2020hololens} \\
    \bottomrule
    \end{tabularx}%
  \label{tab:patient}%
\end{table*}%


\section{Registration and object tracking with the HoloLens}
\label{sec:reg}

\begin{figure}
    \centering
    \includegraphics[width=\columnwidth]{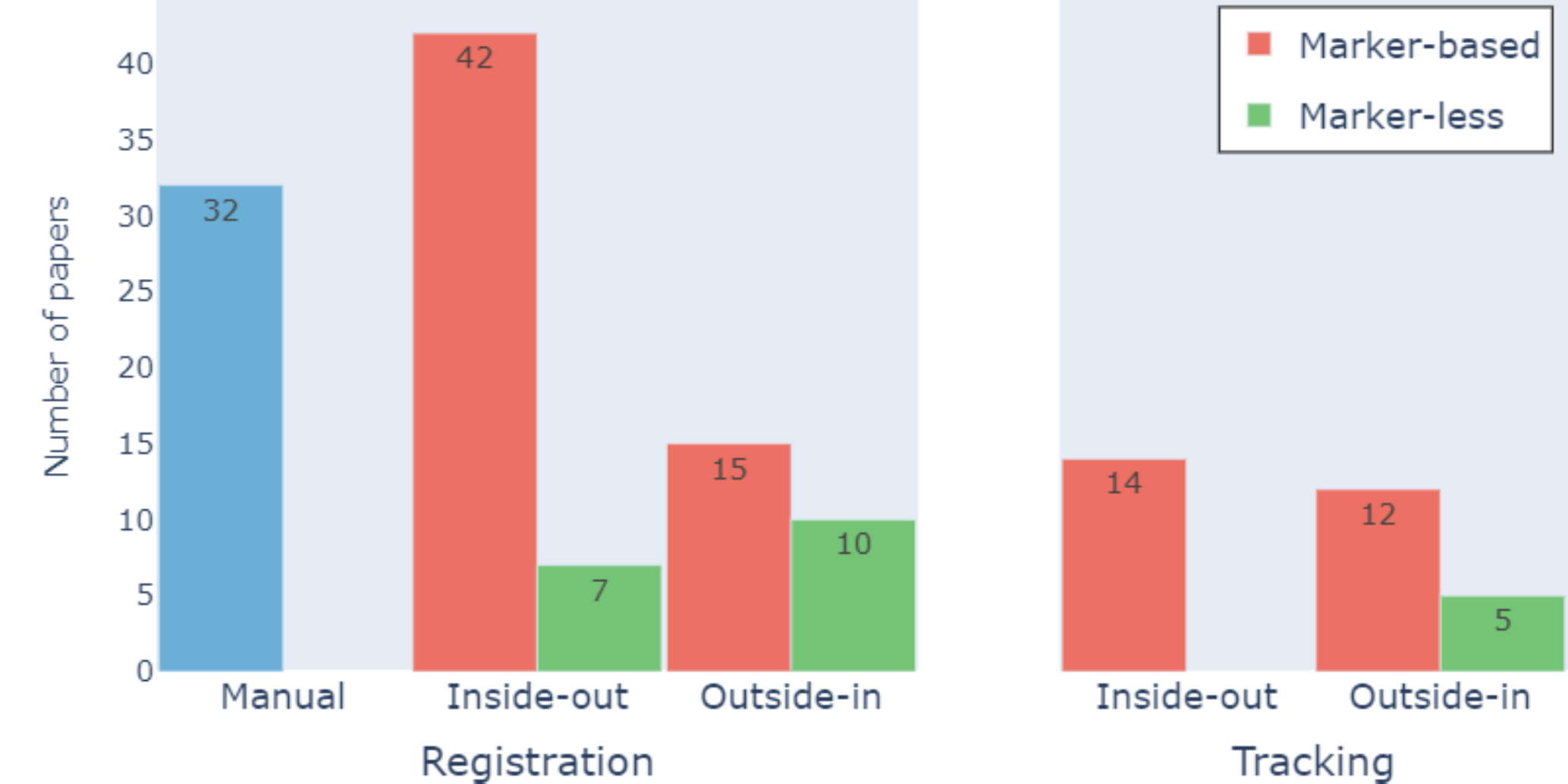}
    \caption{Frequency of registration and tracking methods employed by the reviewed studies. Most works rely on inside-out, marker-based tracking, followed by manual alignment.}
    \label{fig:reg_plot}
\end{figure}

As already mentioned, the registration of virtual content to the physical situation is one of the fundamental concepts of medical AR. Registration permits X-ray visualization for IGI or the localization of the patient's anatomy for surgical navigation, enables hybrid simulators or annotated anatomical specimens in educational settings.

As previously mentioned, tracking for AR mostly focuses on the self-localization of the AR device. Since the HoloLens already provides SLAM, applications in surgical navigation or advanced medical simulators need only track additional non-stationary objects (i.e., medical tools and instruments) with high precision. SLAM is not suitable for this type of dynamic tracking; therefore, other methods need to be implemented.  Registration and tracking are usually closely related, as the same paradigms and methods can be applied to both tasks. In our analysis, we found 99 studies which establish a registration between virtual and real content, which are listed in~\autoref{tab:reg}. 31 studies further integrate methods for object tracking with their AR systems, which are shown in~\autoref{tab:track}. \autoref{fig:reg_plot} visualizes the frequencies of identified paradigms and methods.

\begin{table*}[htb]
  \centering
  \caption{All studies applying registration between physical and virtual content, grouped by registration and tracking paradigm and method.}
    \begin{tabularx}{\textwidth}{p{0.12\textwidth} p{0.15\textwidth} p{0.66\textwidth}}
    \toprule
    Paradigm & Method & Studies \\
    \midrule
        \multirow{2}{*}{Manual (32)} & \makecell[l]{} & 
        \citet{agten2018augmented,frantz2018augmenting,hanna2018augmented,incekara2018clinical,kobayashi2018exploratory,li2018wearable,mcjunkin2018development,pratt2018through,rae2018neurosurgical,gibby2019head,huang2019shared,liu2019percutaneous,lohou2019preliminary,mitsuno2019effective,creighton2020early,fischer2020evaluation,katayama2020intraoperative,nguyen2020augmented,nguyen2020baugmented,nuri2020augmented,saito2020intraoperative,scherl2021augmented,scherl2021baugmented,tian2020validation,buch2021development,gouveia2021breast,gu2021feasibility,iizuka2021potential,ivan2021augmented,meng2021feasibility,schneider2021augmented,sugahara2021mixed}\\
    \midrule
        \multirow{3}{*}{\makecell[l]{Inside-out (52)}} 
        & \makecell[l]{Marker-based (42)} &    \citet{perkins2017mixed,andress2018fly,carbone2018proof,frantz2018augmenting,mahmood2018augmented,moreta2018augmented,qian2018arssist,amini2019augmented,gao2019feasibility,huang2019shared,kalavakonda2019augmented,liebmann2019pedicle,liu2019percutaneous,qian2019aramis,rose2019development,van2019clinical,ferraguti2020augmented,huang2020augmented,jiang2020hololens,kriechling2020augmented,kunz2020infrared,luzon2020value,muller2020augmented,qian2020flexivision,rewkowski2020small,wesselius2021holographic,zuo2020novel,brunzini2021mixed,condino2021hybrid,dennler2021augmented,gu2021feasibility,heinrich2021holopointer,koyachi2021accuracy,li2021augmented,long2021comparison,nguyen2022holous,perez2021effect,qi2021holographic,schneider2021augmented,spirig2021augmented,suzuki2021learning,van2021augmented}\\
        \cmidrule{2-3}
        & \makecell[l]{Marker-less (7)} &  \citet{xie2017holographic,pepe2018pattern,gsaxner2019markerless,pepe2019marker,sylos2019depth,gu2021feasibility,gsaxner2021augmented}\\
    \midrule
        \multirow{3}{*}{\makecell[l]{Outside-in (25)}} 
        & \makecell[l]{Marker-based (15)} &   \citet{condino2018build,el2018augmented,kuzhagaliyev2018augmented,chien2019hololens,de2019hand,fotouhi2019interactive,li2019mixed,meulstee2019toward,rewkowski2020small,ruger2020ultrasound,sun2020fast,glas2021augmented,gu2021feasibility,liu2021wearable,van2021effect}\\
        \cmidrule{2-3}
        & \makecell[l]{Marker-less (10)} &  \citet{kuhlemann2017towards,garcia2018navigation,leuze2018mixed,wu2018augmented,chien2019hololens,wang2019hololens,muangpoon2020augmented,castelan2021augmented,gu2021feasibility}\\
    \bottomrule
    \end{tabularx}%
  \label{tab:reg}%
\end{table*}%

\begin{table*}[htb]
  \centering
  \caption{All studies applying object tracking with the HoloLens, grouped by tracking paradigm and method.}
    \begin{tabularx}{\textwidth}{p{0.12\textwidth} p{0.15\textwidth} p{0.66\textwidth}}
    \toprule
    Paradigm & Method & Studies \\
    \midrule
        \multirow{3}{*}{\makecell[l]{Inside-out (14)}} 
        & \makecell[l]{Marker-based (14)} &    \citet{carbone2018proof,qian2018arssist,gao2019feasibility,liebmann2019pedicle,qian2019aramis,kriechling2020augmented,kunz2020infrared,muller2020augmented,qian2020flexivision,rewkowski2020small,li2021augmented,nguyen2022holous,spirig2021augmented,van2021augmented}\\
    \midrule
        \multirow{3}{*}{\makecell[l]{Outside-in (17)}} 
        & \makecell[l]{Marker-based (12)} &   \citet{el2018augmented,kuzhagaliyev2018augmented,leuze2018mixed,de2019hand,li2019mixed,meulstee2019toward,rewkowski2020small,ruger2020ultrasound,sun2020fast,glas2021augmented,liu2021wearable,van2021effect}\\
        \cmidrule{2-3}
        & \makecell[l]{Marker-less (5)} &  \citet{kuhlemann2017towards,condino2018build,garcia2018navigation,liu2019augmented,muangpoon2020augmented}\\
    \bottomrule
    \end{tabularx}%
  \label{tab:track}%
\end{table*}%

\subsection{Manual registration}
Due to the self-tracking capabilities of the HoloLens, registration between real and virtual content can be achieved simply by manually aligning position, orientation and scale of the virtual items to match their physical counterparts. Since registration is performed for the perspective of the user, factors hindering accurate perception, such as a poor display calibration, may be mitigated. 32 studies in this review adopt such a manual registration technique, mostly by using transformation of objects via on-board input methods (hand gestures and voice commands) or additional input devices, e.g., gamepads~\cite{buch2021development,meng2021feasibility} and keyboards~\cite{nguyen2020augmented}. 

Obviously, manual alignment of virtual content can be time-consuming and ponderous, which affects applicability in clinical settings, where time and personnel are usually scarce.  Landmark-based methods can make manual alignment faster and less cumbersome. They involve the manual annotation of pre-defined anatomical landmarks in the spatial map of the real environment using gestures, which are matched with their virtual counterparts in pre-interventional imaging~\cite{mitsuno2019effective,nguyen2020augmented,nguyen2020baugmented}. However, due to the coarseness of the spatial map and the lack of haptic feedback when selecting landmarks, these approaches may not be reliable or accurate. All manual registration methods have the disadvantage of being static -- if the patient moves, the registration has to be manually adapted accordingly.

\subsection{Inside-out methods}
The built-in sensors of the HoloLens offer several possibilities for inside-out registration and tracking. The advantages of inside-out approaches in medical scenarios are evident: They work in unprepared and unrestricted environments and do not rely on expensive, specialized hardware, thus avoiding extra costs and further cluttering of already densely occupied spaces, such as operation rooms. However, it is still difficult to meet the high demands in accuracy and robustness of medical procedures using inside-out approaches~\cite{sielhorst2008advanced,gsaxner2021archapter}. With 52 occurrences, they are most frequent in our reviewed studies.

\paragraph{Marker-based.}
Marker-based inside-out registration is the most common registration technique identified in this review, employed by 42 studies. Freely available AR libraries, such as Vuforia (PTC Inc, Boston, USA) or ArUco library~\cite{garrido2014automatic}, facilitate optimized, close to real-time detection and tracking of image fiducials via the HoloLens' front-facing RGB camera, which makes marker-based inside-out strategies easy to implement. The most straightforward method for registration, also employed by commercial SNS, is to anchor markers directly to rigid tissue of the patient, e.g., bones. For precisely relating the coordinate frame of the marker to the target anatomy, it is common practice to perform a pre-interventional scan, including the marker. However, attaching markers to patients is invasive and the additional imaging scan may lead to increased radiation exposure of the patient. Additive manufacturing offers an interesting alternative to this route, which allows the creation of patient-specific bone guides or occlusal splints for holding the markers~\cite{moreta2018augmented,gao2019feasibility,koyachi2021accuracy}. In laboratory settings, 3D printing is also commonly used to create custom, marker-embedded phantoms for testing the registration method. Andress et al.~\cite{andress2018fly} even developed a multi-modal marker, allowing intra-interventional marker-based registration. 

Alternatively, landmark-based approaches, where distinct anatomical landmarks are digitized in the coordinate frame of the HoloLens and matched to their virtual counterpart using point based registration, can be used. A marker-tracked pointing device is used for landmark selection in these studies ~\cite{van2019clinical, liebmann2019pedicle, kriechling2020augmented, muller2020augmented, wesselius2021holographic}. To adapt to movements of the patient, a rigidly attached marker is necessary, or the entire procedure has to be repeated.

It is straightforward to extend marker-based inside-out methods for medical instrument tracking by simply attaching markers to the tracked objects as well. 14 reviewed studies apply such a combined strategy. Liu et al.~\cite{liu2021wearable} and Condino et al.~\cite{condino2018build}, combine marker-based, inside-out patient registration with outside-in tracking, using stereo cameras and electromagnetic sensors for tracking, respectively.

A drawback of using fiducial markers is a general lack of robustness and accuracy. It has been shown that the tracking error using common libraries can range from several millimeters to even centimeters~\cite{brand2020accuracy, cao2020image} and is highly dependent on viewing angles, distance, lighting conditions and movement patterns~\cite{leuze2018mixed, jiang2020hololens, luzon2020value, zuo2020novel}. These issues make planar image targets not ideal for highly precise, six degrees of freedom (6DoF) applications, as required in most medical scenarios. In proof-of-concept studies, Kunz et al.~\cite{kunz2020infrared} and Van Gestel et al.~\cite{van2021augmented} have explored the possibility of tracking spherical, IR reflective markers inside-out using the IR sensor of the HoloLens, which appears to be a promising direction.

\paragraph{Marker-less.}
Ten studies explore the possibility of using the various on-board sensors of the HoloLens for inside-out, marker-less registration. An early work by Xie et al.~\cite{xie2017holographic} explored the possibility of surface-based registration of a patient's skin surface with the spatial map created by the HoloLens SLAM. However, the spatial map accessible to developers is very coarse, resulting in insufficiently accurate natural features extractable from it. Hajek et al.~\cite{hajek2018closing} also exploit the HoloLens SLAM by using two devices in a master-worker configuration, while Liu et al.~\cite{liu2019augmented} use image-based matching to align intra-operative X-ray with the patient anatomy. 

Landmark-based registration approaches have been employed as well. For example, Pepe et al.~\cite{pepe2018pattern,pepe2019marker} use automatically detected facial landmarks for registration. From mid 2018 on, the \textit{Research Mode} allowed access to the HoloLens' built in sensors aside from the RGB camera, opening new possibilities for inside-out registration. Sylos-Labini et al.~\cite{sylos2019depth} used automatically detected facial landmarks as well, but showed that, by combining them with the ToF depth data, accuracy can be slightly improved. Gsaxner et al.~\cite{gsaxner2019markerless,gsaxner2021augmented} subsequently introduced a pipeline for fully automatic registration via point cloud matching, using 3D features from ToF depth alone. This method was later also employed by Gu et al.~\cite{gu2021feasibility}, who compared surface-based registration with marker-based and outside-in methods.

\subsection{Outside-in methods}
25 reviewed publications use an outside-in paradigm for registration and tracking. Outside-in approaches rely on external infrastructure for registration and tracking. External infrastructure makes it possible to exploit highly precise, specialized hardware, such as commercial SNS. The high reference accuracy of such systems (usually $\leq$ 1 mm and $\leq$ 1 $^\circ$) makes their integration into an AR environment promising. However, the integration of external systems requires the calibration of coordinate frames between the HoloLens and the navigation device. This procedure usually involves manual and/or semi-automatic steps, which can be cumbersome and disruptive to the clinical workflow, as well as prone to errors and highly subjective~\cite{de2015image}.

\paragraph{Marker-based.}
Most commercially available SNS track passively reflecting markers using stereoscopic IR cameras~\cite{kral2013comparison}. By attaching those markers to the patient, their relative localization in relation to pre-interventional imaging can be determined. The HoloLens can be integrated into such a setup, by affixing markers to the headset as well. Since SNS are designed not only for tracking patients, but, in particular, medical instruments, object tracking can be integrated easily with such systems, and most reviewed studies in this category use this principle. Liu et al.~\cite{liu2021wearable} use stereo cameras and LED markers instead. 

Such marker-based SNS have a high reference precision, often below one millimeter, however, in addition to potential complications resulting from system calibration, they require a constant line-of-sight between IR camera, patient and device, which may restrict movements.

\paragraph{Marker-less.}
Before the HoloLens Research Mode enabled access to the on-board ToF camera of the device, some works integrated external depth sensors with the HoloLens to enable a surface-based registration~\cite{leuze2018mixed,wu2018augmented,chien2019hololens,wang2019hololens}. As an alternative to capture the full surface of patients, again, a sub-set of points in the form of anatomical landmarks can be used, for example, digitized via external electromagnetic trackers~\cite{kuhlemann2017towards,garcia2018navigation,muangpoon2020augmented}. In these scenarios, the electromagnetic sensors have been used for instrument and tool tracking, as well. However, electromagnetic tracking is generally less popular than optical tracking, as it suffers from interference with metallic materials, commonly found in clinical spaces~\cite{kral2013comparison}.


\section{Data and visualization}
\label{sec:viz}

\begin{figure*}[h]
    \centering
    \includegraphics[width=\textwidth]{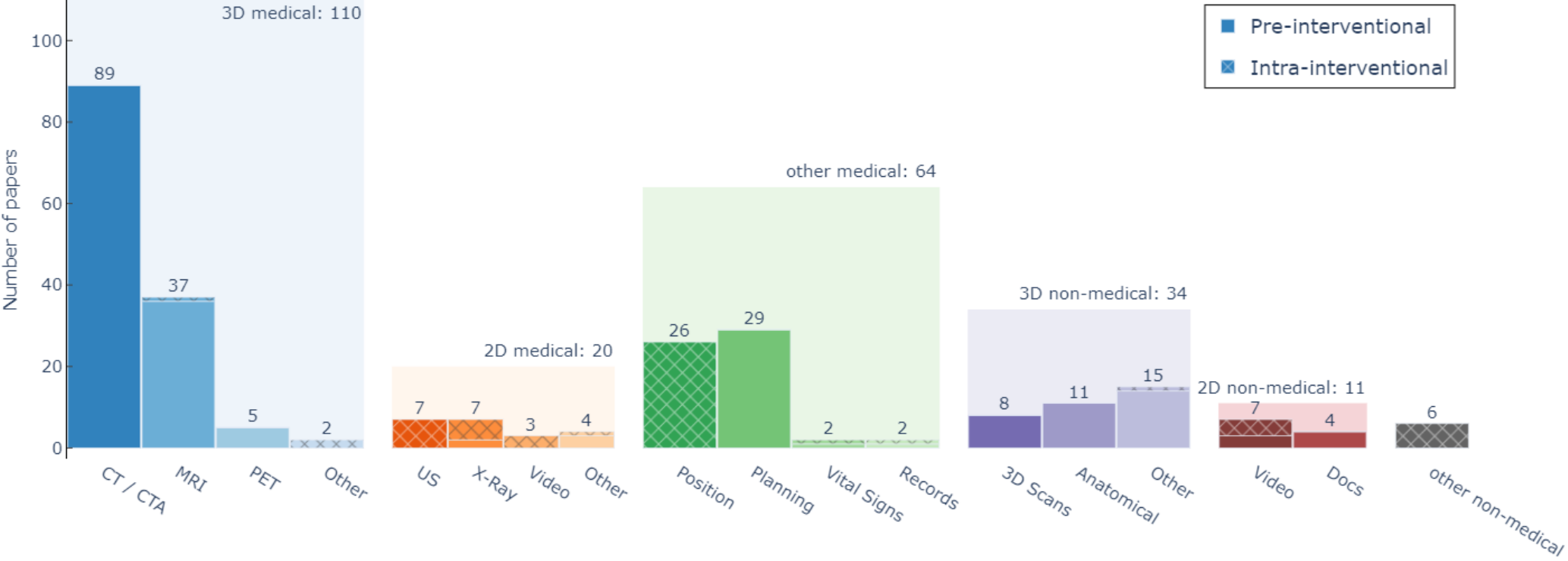}
    \caption{Frequency of data sources used in the reviewed studies. 3D Medical imaging data is, by far, the most common source of data for a visualization in AR.}
    \label{fig:visualization_plot}
\end{figure*}

Various data were visualized in augmented environments through the HoloLens. We distinguish data based on its source (medical or non-medical) and type, according to dimensionality (2D, 3D, other), and discuss how this data is typically visualized. An overview of data source frequencies in the reviewed publications is given in~\autoref{fig:visualization_plot}, and a list of all papers in each category is provided in~\autoref{tab:viz}. Note that most reviewed studies utilize more than one source and type of data -- therefore, multiple mentions are possible.

\subsection{Acquisition time}
Regardless of the source and type, data can further be distinguished based on its acquisition time: Pre-interventional data is acquired offline, processed and uploaded to the HoloLens before the actual intervention. This method allows more complex workflows, including manual manipulations of data. With overall 206 examples, pre-interventional data makes up the majority of sources. 

Intra-operative data is collected at run-time and streamed to the device for visualization. Obviously, intra-operative approaches are technically more complex, since they require a connection between the HoloLens and the raw data source, and necessary processing steps need to be performed automatically, in real-time. Overall, 58 intra-interventional data sources have been identified for this review.

\subsection{Medical data}
\paragraph{3D volumetric medical image data.}
For the majority (110) of reviewed papers, 3D medical images, acquired primarily through CT/CTA (89) and MRI (38), are a main source of data. They are represented as volumetric grids, where each voxel represents a specific value calculated by the imaging device. For visualization, they have to be rendered to present them on the HoloLens display.

Volumetric medical data is conventionally visualized in 2D on monitors in clinical practice, in the form of orthogonal slices through the image volume (mainly axial, sagittal and coronal planes or, sometimes, oblique reformats, so called multi-planar reformations). Since physicians are accustomed to this type of visualization, slice rendering of volumetric data has also been employed in 29 reviewed medical HoloLens systems.

This technique has, of course, the drawback that data is only shown in selected planes. Given a stereoscopic AR display, true 3D visualization is becoming more widely used, mostly in the form of 3D surface renderings, which is computationally efficient and natively supported by all graphics engines compatible with the HoloLens. Furthermore, colors and opacities can easily be modified, enabling visualization techniques such as wire frames or outline visualizations. However, for surface rendering, tissue has to be segmented and converted to polygons prior to visualization, leading to more time intensive workflows and quantization inaccuracies. In contrast, direct volume rendering offers superior image quality~\cite{kutter2008real,kilgus2015mobile} and does not require surface extraction before visualization. Instead, color and opacity are directly computed from the underlying voxel values using specialized transfer functions. Still, performance requirements of volume rendering cannot be easily addressed with mobile hardware, such as the HoloLens. Consequently, only six reviewed studies attempt volume rendering~\cite{frohlich2018holographic,witowski2019augmented,house2020use,ivan2021augmented,gehrsitz2021cinematic,allison2020breast3d}. 

Since data acquisition and reconstruction of 3D volumetric data is relatively costly, only few applications with intra-operative acquisition times exist. Velazco-Garcia et al.~\cite{velazco2021modular} describe a framework for live interactions with MRI scanners. Qian et al.~\cite{qian2019aramis} stream 3D endoscopy data to the HoloLens in real-time, while Southworth et al.~\cite{southworth2020performance} display live 3D cardiac electrophysiology data with the HoloLens.

\paragraph{2D medical image data.}
20 reviewed studies use 2D medical imaging as a data source. Common modalities include X-Ray/fluoroscopy scans (7), ultrasound (7) or endoscopic video (3). Contrary to 3D imaging, 2D modalities usually have short acquisition times (close to or even meeting real-time requirements) and are comparably easy to deploy, and are therefore popular for intra-interventional guidance of procedures. 15 publications in this category support intra-interventional data acquisition during the run-time of the HoloLens.

Analogous to the ordinary clinical practice, 2D imaging data in AR is often visualized on virtual (AR) monitors, which can be anchored to the head gaze of the HoloLens wearer. Another possibility is to position 2D images on 3D planes in the environment, which allows an in-situ visualization, if a registration between imaging data and patient is available.

\paragraph{Other data from medical sources.}
In many situations, it is beneficial to integrate other medical data not stemming from medical imaging into the workflow. Medical planning data is a particularly common example, with 29 publications integrating planning data into their workflows. This data is usually created manually and pre-operatively by medical professionals before an intervention on the basis of medical imaging. It can include access points, tool trajectories, cutting lines, resection margins and target positions of implants, amongst others. This type of data is usually translated into geometric primitives, which are displayed in relation to the target anatomy. 

For intra-interventional data, the positional coordinates of medical tools (such as needles, wires, or screws) or other tracked objects (parts of the anatomy, imaging systems) obtained from outside-in or inside-out navigation systems are the most common data source. Mostly, these objects are represented by geometric primitives or 3D models, which are transformed according to the positional information. However, a simple numerical representation is also used in some studies~\cite{liebmann2019pedicle,liu2019augmented,kriechling2020augmented}. 

Other medical data sources, which have been captured both pre- and intra-interventionally, include vital signs or other biosignals and patient records, which can be displayed on virtual monitors in AR. 

\subsection{Non-medical data}
\paragraph{3D data.}
The inherent ability of the HoloLens for stereoscopic rendering make all sorts of 3D meshes an obvious choice of data source for AR visualizations.

Eight studies use 3D scans of patients, captured with depth or stereo cameras, instead of volumetric medical imaging, mostly for the purpose of image-to-patient registration.

Contrary to medical 3D data, such scans can only capture the surface of patients and do not inform about the underlying anatomy. In particular in educational scenarios (targeting both patients and students), the visualization of anatomical models, created by medical artists, is common and used in 11 studies. Both of these data sources have exclusively been deployed pre-interventionally to the HoloLens.

Other non-medical 3D data include geometric primitives and non-anatomical models, e.g., for creating serious games~\cite{aruanno2017hololens,garzotto2018hololearn,aruanno2019memholo} or procedure simulations~\cite{wang2017augmented,cecil2018design,maniam2020exploration,rohrbach2019augmented,velazco2021modular}.

\paragraph{2D data.}
A small number of eleven studies visualize non-medical two-dimensional data in the form of pre-recorded or live streamed videos or documents. As with 2D medical data, it is usually displayed on virtual monitors anchored to the display or environment.

\paragraph{Other data.}
Six works have explored the possibility of integrating other data, in most cases coming from the HoloLens itself, into their applications. Three publications track the user wearing the HoloLens, to measure gait parameters~\cite{geerse2020quantifying,koop2020hololens} or guide the user~\cite{yamashita2017pedestrian}. Two publications utilize the gaze data from the HoloLens~\cite{sun2019validity,heinrich2021holopointer}. Only Sharma et al.~\cite{sharma2018mixed} use an external data source, namely IMU data, for training limb prosthesis control. 

\makeatletter%
\if@twocolumn%
  \afterpage{\afterpage{\afterpage{
\onecolumn
\topcaption{All studies grouped by data source and type, data modality and acquisition time.}
\tablefirsthead{%
\toprule
\makecell[l]{Source \\ \& Type} & \makecell[l]{Modality} & \makecell[l]{Acquisition \\ Time} & \makecell[l]{Studies} \\
\midrule}
\tablehead{%
\toprule
\makecell[l]{Source \\ \& Type} & \makecell[l]{Modality} & \makecell[l]{Acquisition \\ time} & \makecell[l]{Studies} \\
\midrule}
%
\begin{xtabular*}{\linewidth}{m{0.06\textwidth} m{0.08\textwidth} m{0.09\textwidth} m{0.67\textwidth}}
    \label{tab:viz}\\
    \shrinkheight{-3\normalbaselineskip}
    \multirow{5}[7]{*}{\begin{sideways}3D medical (110)\end{sideways}} 
        & \makecell[l]{CT \& \\ CTA \\ (89)} &  Pre (89)   &
        \citet{bucioli2017holographic,kuhlemann2017towards,sauer2017mixed,agten2018augmented,condino2018build,el2018augmented,frantz2018augmenting,frohlich2018holographic,garcia2018navigation,hajek2018closing,kobayashi2018exploratory,li2018wearable,mahmood2018augmented,mcjunkin2018development,moreta2018augmented,pratt2018through,rae2018neurosurgical,wu2018augmented,affolter2019applying,brun2019mixed,checcucci20193d,de2019hand,gao2019feasibility,gibby2019head,gsaxner2019markerless,huang2019shared,kalavakonda2019augmented,kubben2019feasibility,li2019mixed,liebmann2019pedicle,liu2019augmented,liu2019percutaneous,lohou2019preliminary,maniam2020exploration,mitsuno2019effective,mitsuno2019telementoring,moosburner2019real,rose2019development,sylos2019depth,van2019clinical,wake2019patient,wang2019hololens,witowski2019augmented,al2020effectiveness,allison2020breast3d,bulliard2020preliminary,cartucho2020multimodal,creighton2020early,ferraguti2020augmented,fischer2020evaluation,fitski2020mri,galati2020experimental,jiang2020hololens,katayama2020intraoperative,kriechling2020augmented,kumar2020use,luzon2020value,muangpoon2020augmented,muller2020augmented,nguyen2020augmented,nguyen2020baugmented,perkins2020patient,saito2020intraoperative,sun2020fast,tian2020validation,zuo2020novel,buch2021development,cofano2021augmented,condino2021hybrid,dennler2021augmented,dennler2021baugmented,gehrsitz2021cinematic,glas2021augmented,gsaxner2021augmented,gu2021feasibility,iizuka2021potential,koyachi2021accuracy,li2021augmented,long2021comparison,meng2021feasibility,qi2021holographic,saito2022intraoperative,schneider2021augmented,spirig2021augmented,sugahara2021mixed,suzuki2021learning,van2021augmented,wake2021workflow,wesselius2021holographic} \\ %
        \cmidrule{2-4}          
        & \multirow{2}[2]{*}{\makecell[l]{MRI \\ (37)}} & Pre (35)   & \citet{mojica2017holographic,perkins2017mixed,xie2017holographic,carbone2018proof,frohlich2018holographic,incekara2018clinical,jang2018three,leuze2018mixed,gsaxner2019markerless,kubben2019feasibility,soulami2019mixed,van2019clinical,wake2019patient,allison2020breast3d,cartucho2020multimodal,ferraguti2020augmented,galati2020experimental,gnanasegaram2020evaluating,house2020use,kumar2020use,nguyen2020augmented,nguyen2020baugmented,pelanis2020use,tian2020validation,condino2021hybrid,gehrsitz2021cinematic,gsaxner2021augmented,iizuka2021potential,ivan2021augmented,morales2021holographic,qi2021holographic,scherl2021augmented,scherl2021baugmented,van2021augmented,van2021effect,wake2021workflow} \\
            \cmidrule{3-4}
            &       & Intra (2) & \citet{velazco2021evaluation,velazco2021modular} \\
        \cmidrule{2-4}          
        & PET (5)   & Pre (5)   & \citet{frohlich2018holographic,pepe2018pattern,pepe2019marker,gsaxner2019markerless,gsaxner2021augmented,galati2020experimental} \\
        \cmidrule{2-4}          
        & Other (2) & Intra (2) & \citet{qian2019aramis,southworth2020performance} \\
    \midrule
    \multirow{6}[7]{*}{\begin{sideways}2D medical (20)\end{sideways}} 
        & US (7)    & Intra (7) & \citet{el2018augmented,garcia2018navigation,kuzhagaliyev2018augmented,mahmood2018augmented,cartucho2020multimodal,ruger2020ultrasound,nguyen2022holous} \\
        \cmidrule{2-4}          
        & \multirow{2}[2]{*}{\makecell[l]{X-Ray \\ (7)}} & Pre (2)  & \citet{qian2017comparison,galati2020experimental} \\
            \cmidrule{3-4}
            &       & Intra (5) & \citet{andress2018fly,deib2018image,fotouhi2019interactive,liu2019augmented,al2020effectiveness} \\
        \cmidrule{2-4}          
        & Video (3) & Intra (3) & \citet{qian2018arssist,al2020effectiveness,qian2020flexivision} \\
        \cmidrule{2-4}          
        & \multirow{2}[2]{*}{Other (4)} & Pre (3)  & \citet{hanna2018augmented,huang2020augmented,robinson2020evaluating} \\
            \cmidrule{3-4}
            &       & Intra (1) & \citet{cui2017augmented} \\
    \midrule
    \multirow{6}[8]{*}{\begin{sideways}other medical (64)\end{sideways}} 
        & Position (26) & Intra (26) & \citet{kuhlemann2017towards,andress2018fly,carbone2018proof,condino2018build,garcia2018navigation,hajek2018closing,kuzhagaliyev2018augmented,qian2018arssist,fotouhi2019interactive,gao2019feasibility,li2019mixed,liebmann2019pedicle,liu2019percutaneous,meulstee2019toward,qian2019aramis,kriechling2020augmented,muangpoon2020augmented,muller2020augmented,qian2020flexivision,iqbal2021augmented,li2021augmented,liu2021wearable,spirig2021augmented,van2021augmented,van2021effect,velazco2021evaluation} \\
        \cmidrule{2-4}          
        & Planning (29) & Pre (29) & \citet{carbone2018proof,condino2018build,kobayashi2018exploratory,li2018wearable,pratt2018through,gibby2019head,li2019mixed,liebmann2019pedicle,liu2019percutaneous,lohou2019preliminary,wang2019hololens,ferraguti2020augmented,kriechling2020augmented,muller2020augmented,maniam2020exploration,tian2020validation,cofano2021augmented,condino2021hybrid,dennler2021baugmented,glas2021augmented,li2021augmented,liu2021wearable,long2021comparison,meng2021feasibility,morales2021holographic,spirig2021augmented,suzuki2021learning,van2021augmented,wesselius2021holographic} \\
        \cmidrule{2-4}          
        & \multirow{2}[2]{*}{\makecell[l]{Vital \\ Signs (2)}} & Pre (1)  & \citet{qian2017comparison} \\
            \cmidrule{3-4}
            &       & Intra (1) & \citet{sirilak2018new} \\
        \cmidrule{2-4}          
        & \multirow{2}[2]{*}{\makecell[l]{Records \\ (2)}} & Pre (1)  & \citet{perkins2020patient} \\
            \cmidrule{3-4}
            &       & Intra (1) & \citet{deib2018image} \\
    \midrule
    \multirow{4}[6]{*}{\begin{sideways}\makecell{3D non-medical \\ (34)}\end{sideways}} 
        & 3D Scans (8) & Pre (8)  & \citet{hanna2018augmented,amini2019augmented,chien2019hololens,talaat2019three,nuri2020augmented,proniewska2020three,kumar2021novel,liu2021wearable} \\
        \cmidrule{2-4}          
        & \makecell[l]{Anatomical \\ model (11)} & Pre (11)  & \citet{balian2019feasibility,stojanovska2019mixed,antoniou2020biosensor,moro2021hololens,robinson2020evaluating,rositi2021presentation,ruthberg2020mixed,bogomolova2021development,brunzini2021mixed,castelan2021augmented,putnam2021virtual} \\
        \cmidrule{2-4}
        & \multirow{2}[2]{*}{Other (15)} & Pre (14) & \citet{aruanno2017hololens,wang2017augmented,cecil2018design,garzotto2018hololearn,sharma2018mixed,aruanno2019memholo,maniam2020exploration,meulstee2019toward,palermo2019augmented,rohrbach2019augmented,geerse2020quantifying,janssen2020effects,rewkowski2020small,velazco2021evaluation} \\
            \cmidrule{3-4}
            &       & Intra (1) & \citet{angelopoulos2019enhanced} \\
    \midrule    
    \multirow{3}[4]{*}{\makecell[l]{2D non-\\medical \\(11)}} 
        & \multirow{2}[2]{*}{Video (7)} & Pre (3) & \citet{sun2019validity,lu2020integrating,schoeb2020mixed} \\
            \cmidrule{3-4}
            &       & Intra (4)& \citet{sirilak2018new,mitsuno2019telementoring,glick2020augmenting,cofano2021augmented} \\
        \cmidrule{2-4}
        & Docs (4)  & Pre (4)& \citet{hanna2018augmented,sirilak2018new,galati2020experimental,rositi2021presentation} \\
    \midrule
    other non-medical (6) & -     & Intra (6) & \citet{yamashita2017pedestrian,sharma2018mixed,sun2019validity,geerse2020quantifying,koop2020hololens,heinrich2021holopointer} \\
    \bottomrule
\end{xtabular*}
\twocolumn
}}}%
\else
  \input{tables/viz_table_xtab}%
\fi
\makeatother


\section{Evaluation of medical HoloLens applications}
\label{sec:eval}

In general, an objective evaluation of AR applications is challenging, because each user has a different perception of augmented content, depending on individual anatomy (interpupillary distance, eye sight), familiarity with the technology, familiarity with 3D visualizations in general~\cite{rosser2007impact} and external influences, such as comfort while wearing the AR device. In comparison to other areas of computer science, no benchmarks, datasets or standard protocols exist to evaluate AR experiences and usefulness. Clinically, comparative clinical trials, measuring and comparing parameters about treatment outcomes, such as treatment time, number and severity of complications or survival rate, are considered gold standard. However, each AR application requires approval by a relevant agency or committee before it can be tested on cadavers, healthy human subjects or even patients. Depending on the executive research institution and national regulations, obtaining such an approval and the quantitative data that comes with it, can be very difficult for researchers. Therefore, in our reviewed studies, a large variety of evaluation metrics have been collected in distinct experimental scenarios, which are summarized in~\autoref{tab:eval}.

\begin{figure}[hb]
    \centering
    \includegraphics[width=\columnwidth]{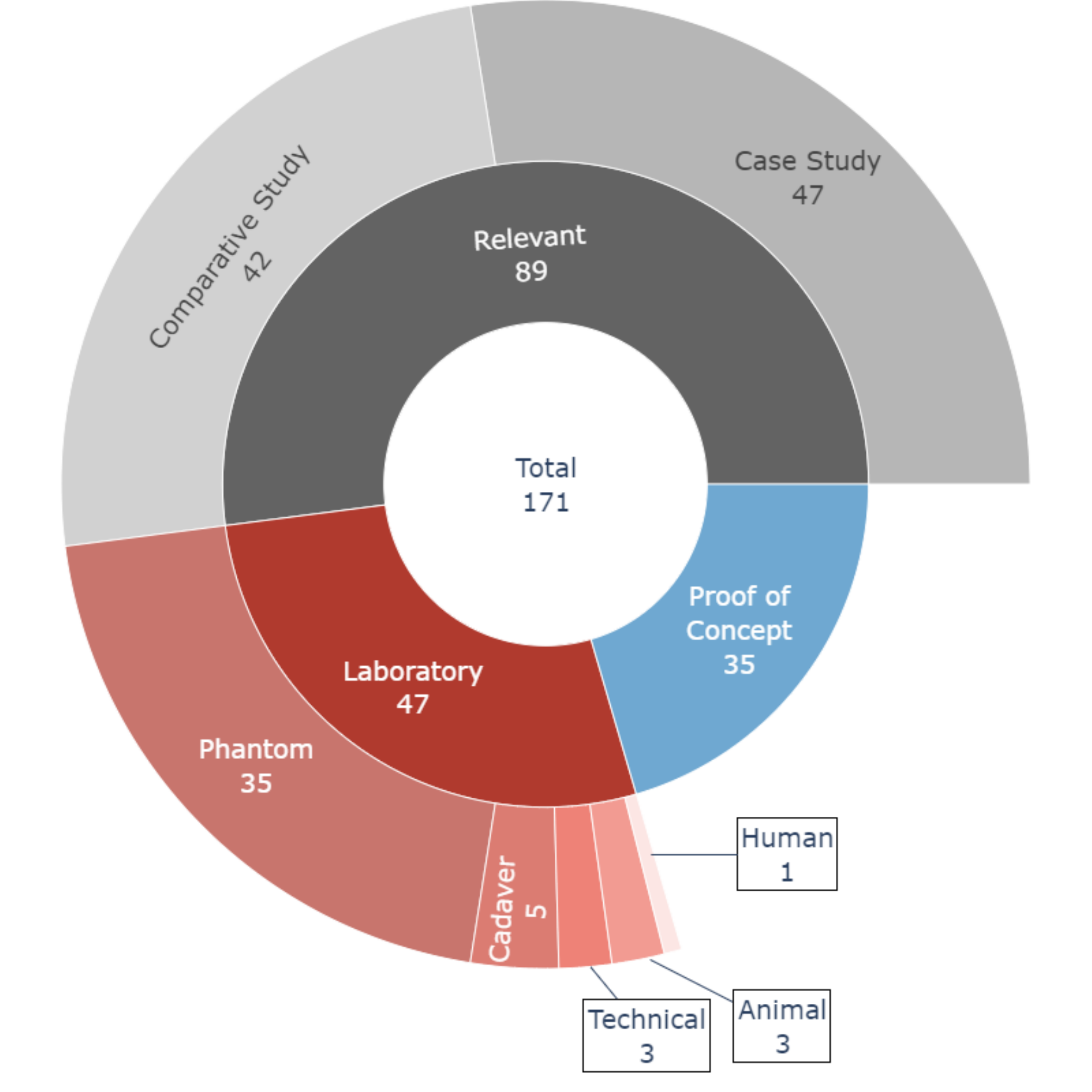}
    \caption{Number of papers for each experimental setting (inner circle) and experimental level (outer circle).}
    \label{fig:eval_plot}
\end{figure}

\subsection{Evaluation scenario}

We first analyze the reviewed publications with regard to the evaluation scenario. Inspired by the Technology Readiness Level~\cite{mankins1995technology}, we group the studies according to their evaluation settings, which is shown in~\autoref{fig:eval_plot}:

\paragraph{Proof of concept studies }
focus on reporting a medical problem, how AR could overcome it and describe their prototype workflows and applications. Sometimes, anecdotal or informal feedback from users or general observations are reported, but, in general, these studies do not follow a rigorous experimental protocol and do not collect quantitative or qualitative measurements. Therefore, it is difficult to draw general conclusions from them. With 35 papers, proof of concept studies are in the minority.

\paragraph{Laboratory studies} typically focus on the technical aspects of their applications and report quantitative measurements. We identify 47 records in this category. The study can be carried out using only hardware (e.g., the HoloLens), or on cadavers, animals or humans (healthy or patients). Most commonly, however, phantoms are used to collect measurements. Specialized medical phantoms, which include realistic anatomical structures and tissue characteristics, are commercially available, however, they are very expensive. Consequently, many researchers resort to additive manufacturing (i.e., 3D printing) to replicate the target anatomy or build more abstract phantoms.

\paragraph{Studies performed in a relevant environment} evaluate their AR systems directly in the environment in which it should be implemented. Such an approach involves the usage of the system by one or more individuals of the intended target group -- either clinicians, patients or medical students. Most of the time, qualitative feedback in the form of questionnaires is collected from them, although quantitative measurements, for example measuring task performance, might also be taken. As seen in~\autoref{fig:eval_plot}, most of the reviewed studies (89) fall into this category, which indicates advanced research maturity. We further distinguish between non-comparative studies, where results acquired through AR are not compared to another method (for example, case series or uncontrolled cohort studies), and comparative studies, which provide comparisons to non-AR conditions. The latter are most conclusive about the possible advantages and implications of the HoloLens in their domain. 

\subsection{Quantitative metrics}
Quantitative metrics are often focused on technical aspects of the AR system. Therefore, acquiring them does not require a large number of test subjects and, instead, can be done by individuals. However, they can also characterize the performance of individuals in carrying out certain tasks. In this case, quantitative measures are usually closely related to the application scenario.

\paragraph{Technical performance metrics.}
Several works, in particular in the area of data display, measure performance metrics of the HoloLens itself, such as hardware utilization, frame rate, power usage, execution time and latency. The studies come to the conclusion that the HoloLens is suitable for displaying pre- and intra-interventional medical data given an appropriate software framework, also within safety critical environments, such as the OR. A commonly reported limiting factor is battery life, which restricts device usage to around two hours, which is too short for many medical interventions.

The HoloLens was also compared to other OST-HMD devices for medical usage. Qian et al.~\cite{qian2017comparison} evaluated the HoloLens, Epson Moverio BT-200 and ODG R-7 for displaying object-anchored 2D medical data, and concluded that the HoloLens is the best choice in terms of contrast, frame rate and perceived task load. Moosburner et al.~\cite{moosburner2019real} compare the HoloLens to the Meta 2 (Meta Company, San Mateo, California,  USA) and found that, albeit the HoloLens was criticized for having a comparably small FoV and being more complicated and difficult to operate, medical students preferred it over the competitor, as it does not rely on a wired connection to a powerful external computer and presented virtual models more stably. 

\paragraph{Accuracy metrics.}

For registration and tracking, accuracy metrics, measuring the spatial distance between the virtual and real position of an object, are usually acquired. 

While many different measures can be computed, the target registration error (TRE) is one of the most commonly and consistently used metrics for evaluating registration accuracy, and has been employed by 22 reviewed studies.

TRE measures the Euclidean distance between 3D target points in the physical world and their virtual counterparts. Studies evaluating the TRE report averages of just above 1 mm and up to 40 mm. E.g., for a registration using outside-in tracking, El Hariri et al.~\cite{el2018augmented} report a TRE of 36.9 mm, while Kuhlemann et al.~\cite{kuhlemann2017towards}, Li et al.~\cite{li2019mixed} and Sun et al.~\cite{sun2020fast} report much lower values of 4.3 mm, 2.2 mm and 1.3 mm, respectively. For manual registration, the reported error spectrum is also large, ranging between 20 mm~\cite{buch2021development} and 3 mm~\cite{tian2020validation}. Registration using image fiducials seems to be the most reliable in terms of TRE, with values in the 2 mm region~\cite{condino2018build,moreta2018augmented}, but several studies show that the achievable accuracy with image fiducials is highly dependent on lumination, viewing angle and movement~\cite{jiang2020hololens,luzon2020value,zuo2020novel}. Whether the reported registration accuracies are acceptable is, of course, contingent upon the clinical scenario. However, most studies express the need to reduce the registration error before clinical usability. While TRE provides some comparability between registration methods, measuring it involves the selection or digitization of matching landmark points, which is itself a subjective, error-prone procedure, encumbered by a lack of haptic feedback, fine-grained input possibilities and depth perception. These problems explain the large variability reported for this metric. 

A variant of the TRE is the fiducial registration error (FRE), which uses fiducials used in landmark-based registration approaches as target points for error computation. Three publications report FRE. For example, Van Doormaal et al.~\cite{van2019clinical} compared FRE achievable with marker-based inside-out registration using landmarks with a conventional SNS registration. They found that the AR system is less accurate and not yet suitable for clinical application. However, it has been shown that FRE does not correlate with the TRE and thus, does not inform much about the actual registration accuracy~\cite{fitzpatrick2009fiducial}. 

Another common measure, evaluated in 18 studies, is the target visualization error (TVE), which measures the re-projection error between physical and virtual objects as perceived by the user, e.g., using a ruler or a millimeter grid or by marking the virtual projection directly on the real counterpart. Most studies report TVE values in the millimeter region. Three studies compare the TVE achieved using the HoloLens with a non-AR baseline: Ivan et al.~\cite{ivan2021augmented} found no significant difference to a commercial SNS in terms of TVE. Qi et al.~\cite{qi2021holographic} state that AR could reach the reference precision in 80\% of cases, while Incekara et al.~\cite{incekara2018clinical} determined that only in 38\% of cases the reference could be met, and the mean deviation of 4 mm between HoloLens and SNS is too large for clinical applicability. Still, the manual measurement of TVE is, again, subject to operator bias.

Registration error, measured in three studies, calculates the deviation between the source-to-target transformation computed by the employed algorithm and a reference transformation obtained from a reference tracking system. Analogously, tracking error compares the pose of a tracked object to a ground truth, ideally in six degrees of freedom. Since the reference system and the HoloLens need to be calibrated, such experiments are complicated to set up. Therefore, many studies report a simplified tracking error, e.g., in 2D~\cite{liu2019augmented} or positional only~\cite{kunz2020infrared,liu2021wearable}.

In clinical interventions where pre-interventional planning data is available, the target deviation error (TDE), which measures the Euclidean distance between a pre-operatively planned target point and the actual point after intervention, can be determined. A typical scenario is the insertion of objects, such as needles, wires or screws, into a phantom, cadaver or patient under AR guidance. 
After insertion, post-operative imaging is acquired, which can be compared to the planning. This type of clinically specific evaluation is most objective and informative about how an AR system can support the intervention in question. We identified 24 publications evaluating TDE.
Several studies perform such a clinically specific evaluation by comparing the outcome of an AR-supported procedure to a non-AR control condition; however, results are inconclusive. Several studies~\cite{agten2018augmented,andress2018fly,liu2019percutaneous,muller2020augmented,long2021comparison} compare needle/wire placements under AR guidance with a conventional, flouroscopy-guided procedure. They found that placements in AR were slightly less accurate than in the reference condition, although AR guidance lead to faster task completion. Andress et al.~\cite{andress2018fly} and Long et al.~\cite{long2021comparison} further point out that, with AR, less radiation was required during image-guided procedures. Li et al.~\cite{li2019mixed}, Ruger et al.~\cite{ruger2020ultrasound} and Glas et al.~\cite{glas2021augmented}, report favorable needle insertion accuracies in AR-guided procedures versus conventional image guided procedures. Compared to freehand, non-guided procedures, AR could improve both accuracy and number of successful task completions in placement tasks~\cite{ferraguti2020augmented,dennler2021augmented,van2021effect}.

\paragraph{Task-specific scores.}
Studies using the HoloLens for supporting specific medical tasks usually report some quantification of task completion. The task completion time (TCT) is most commonly measured, namely in 32 reviewed studies. Most comparative studies report that AR guidance helped users in carrying out tasks faster~\cite{galati2020experimental,agten2018augmented,al2020effectiveness,andress2018fly,liu2019percutaneous,muller2020augmented,long2021comparison,heinrich2021holopointer,glas2021augmented,sharma2018mixed,ferraguti2020augmented,qian2020flexivision,suzuki2021learning}, while others did not report significant differences~\cite{wang2017augmented,deib2018image}. Only Qi et al.~\cite{qi2021holographic} and Rohrbach et al.~\cite{rohrbach2019augmented} report longer TCT for the AR condition, however, the latter application is targeted as Alzheimer's patients, who may have more difficulties in adapting to novel technology, such as AR and the HoloLens.

The number of successful task completions (NSC) is measured in 12 studies. Most studies report favorable outcomes of HoloLens usage in terms of NSC ~\cite{dennler2021augmented,schneider2021augmented,sharma2018mixed,qian2020flexivision}. Only Agten et al.~\cite{agten2018augmented} found that AR actually leads to less successful outcomes, compared to a conventional image-guided procedure.

The effectiveness of AR for learning in an educational scenario can be quantitatively measured by comparing exam scores between AR-supported learners and a control group. Seven studies perform such an evaluation. No statistically significant knowledge improvement was found between students receiving AR lectures through the HoloLens versus students undergoing conventional anatomy courses based on cadavers~\cite{stojanovska2019mixed,robinson2020evaluating,ruthberg2020mixed}. Robinson et al.~\cite{robinson2020evaluating}, however, highlight that students perceived the AR activity more favorably. Similar findings are described in comparison to other computerized learning methods by \cite{antoniou2020biosensor,gnanasegaram2020evaluating,moro2021hololens} -- while student engagement, motivation and excitement is typically higher for HoloLens-based education, the outcomes in terms of learning effect are not significantly different. 

\subsection{Qualitative metrics}
We define qualitative metrics as parameters and data, which reflect the personal opinion of individuals, and can, therefore, not be objectively and repeatably measured. Usually, they are collected from application users by the means of questionnaires or interviews. Since AR experiences are highly individual, qualitative metrics can be considered equally if not more important than quantitative measures. After all, theoretical benefits of medical AR are negligible if the system that delivers them is deemed cumbersome or fails to meet the user's needs.

Commonly, questionnaires use a Likert scale, where respondents express their level of agreement or disagreement with certain statements. 42 reviewed publications use such questionnaires for evaluating various system aspects. Examples include general comfort, image quality and audio quality of the HoloLens and its suitability for medical applications~\cite{condino2018build,jang2018three,sirilak2018new,moosburner2019real,galati2020experimental,kumar2020use,al2020effectiveness,scherl2021augmented,dennler2021augmented}, the effectiveness of certain types of visualization~\cite{brun2019mixed,wake2019patient,house2020use,gehrsitz2021cinematic} or, most commonly, how well the proposed application can support a certain procedure. 

Generally, the reported questionnaire outcomes are favorable towards the HoloLens and AR, and the common consensus is that AR can have a large impact in the medical domain. However, limitations of the device itself, such as the small field of view, short battery life and relative discomfort while wearing it are frequently mentioned. For IGI or navigation applications, users also frequently noticed a lack of registration accuracy or a drift of virtual content due to instabilities in the HoloLens SLAM, which negatively influenced user ratings. In these scenarios, issues of depth perception, where users perceived internal anatomy to be on top of, not within, the patient, were also frequently mentioned. 

Some reviewed studies employed standardized questionnaires, with the NASA Task Load Index (NASA-TLX), a tool to assess subjective workload, being the most commonly used one.
A drawback of the NASA-TLX is that it is only fully descriptive in comparative studies, where it is measured for several conditions. A classification or interpretation of a single final score is, generally, not substantial. Unfortunately, only a few comparative studies measure the NASA-TLX -- two of them report a reduced task load for AR-supported procedures~\cite{ferraguti2020augmented,qian2020flexivision}, Rüger et al.~\cite{ruger2020ultrasound} found no significant difference and Saito et al.~\cite{saito2020intraoperative} found that the mental demand was higher for participants using AR. 

The System Usability Scale (SUS)~\cite{brooke1996sus} was applied in five studies as a measure for application usability. Compared to the NASA-TLX, the advantage of SUS is that it is a fast way of classifying the ease of use of a system, even without a comparison. Generally, overall scores greater than 68 are considered above average; furthermore, an adjective rating scale has been proposed~\cite{bangor2009determining}. Only two reviewed studies compute the overall SUS, both reporting above average usability with SUS values of 71.5~\cite{amini2019augmented} and 74.8~\cite{gsaxner2021augmented}, scoring a "Good" on the adjective rating scale. While these ratings are encouraging, they suggest that there is room for improvement.

\makeatletter%
\if@twocolumn%
\afterpage{
\onecolumn
\tablecaption{Summary of evaluation strategies for medical HoloLens applications. Studies are grouped according to their evaluation setting and level. For each study, we report the acquired qualitative (Qual) and quantitative (Quant) measures.}
\tablefirsthead{%
\toprule
\multicolumn{3}{c}{Proof of Concept Studies (no measures reported)} \\
\midrule
\multicolumn{3}{p{\textwidth}}{\citet{bucioli2017holographic,cui2017augmented,sauer2017mixed,xie2017holographic,carbone2018proof,cecil2018design,garzotto2018hololearn,hanna2018augmented,kobayashi2018exploratory,kuzhagaliyev2018augmented,mahmood2018augmented,pratt2018through,affolter2019applying,kalavakonda2019augmented,kubben2019feasibility,lohou2019preliminary,mitsuno2019telementoring,soulami2019mixed,witowski2019augmented,allison2020breast3d,boyd2020augmented,fitski2020mri,katayama2020intraoperative,lu2020integrating,maniam2020exploration,perkins2020patient,proniewska2020three,castelan2021augmented,gouveia2021breast,iizuka2021potential,morales2021holographic,saito2022intraoperative,sugahara2021mixed,wake2021workflow,wesselius2021holographic}} \\
\bottomrule
\toprule
\multicolumn{3}{c}{Laboratory Studies (Quantitative measures)} \\
\midrule
Level & Study & Measures \\
\midrule}
\tablehead{%
\midrule
Level & Study & Measures \\
\midrule}
\begin{xtabular*}{\linewidth}{ p{0.07\textwidth} p{0.24\textwidth}  p{0.61\textwidth}}
\label{tab:eval}\\
\multirow{3}[2]{*}{Technical} 
    & \citet{mojica2017holographic} & CPU usage, memory consumption, FPS \\
    & \citet{frohlich2018holographic} & CPU usage, GPU usage, memory consumption, FPS \\
    & \citet{velazco2021modular} & FPS, Latency \\
\midrule      
\multirow{35}[2]{*}{\begin{sideways}Phantom\end{sideways}} 
    & \citet{agten2018augmented} & Task completion time, Number of successful completions \\
    & \citet{el2018augmented} & Target registration error \\
    & \citet{frantz2018augmenting} & Registration time, Target visualization error, content drift \\
    & \citet{garcia2018navigation} & Latency \\
    & \citet{hajek2018closing} & Calibration error (hand-eye), Target registration error, Number of successful completions \\
    & \citet{leuze2018mixed} & Tracking error, Registration error \\
    & \citet{moreta2018augmented} & Target registration error, Target deviation error (surgical guide) \\
    & \citet{qian2018arssist} & Target visualization error, Tracking error, FPS \\
    & \citet{rae2018neurosurgical} & Target visualization error, Registration time \\
    & \citet{wu2018augmented} & Target registration error, Registration time \\
    & \citet{chien2019hololens} & Target registration error, Registration time \\
    & \citet{de2019hand} & SLAM accuracy, Latency, Target visualization error \\
    & \citet{fotouhi2019interactive} & Calibration error (hand-eye), Tracking accuracy \\
    & \citet{gibby2019head} & Target deviation error (needle), Task completion time \\
    & \citet{gsaxner2019markerless} & Target registration error, Registration error, Registration time \\
    & \citet{huang2019shared} & Target visualization error, Task completion time \\
    & \citet{liebmann2019pedicle} & Fiducial registration error, Registration time, Target deviation error (screw) \\
    & \citet{liu2019augmented} & Tracking error (2D) \\
    & \citet{liu2019percutaneous} & Target deviation error (screw) \\
    \shrinkheight{-18\normalbaselineskip}
    & \citet{meulstee2019toward} & Calibration error (hand-eye), Target deviation error (model)\\
    & \citet{mitsuno2019effective} & Registration time, Target visualization error \\
    & \citet{sylos2019depth} & Target visualization error \\
    & \citet{wang2019hololens} & Target registration error, Registration time \\
    & \citet{creighton2020early} & Target registration error \\
    & \citet{gu2021feasibility} & Registration error \\
    & \citet{huang2020augmented} & Target visualization error \\
    & \citet{jiang2020hololens} & Target registration error \\
    & \citet{kriechling2020augmented} & Target deviation error (k-wire) \\
    & \citet{kunz2020infrared} & Tracking error (relative, position) \\
    & \citet{luzon2020value} & Target registration error \\
    & \citet{nguyen2020augmented} & Target visualization error \\
    & \citet{rewkowski2020small} & Calibration error, Latency \\
    & \citet{sun2020fast} & Target registration error, Registration time \\
    & \citet{perez2021effect} & Target visualization error \\
    & \citet{van2021augmented} & Latency \\
\multirow{3}[2]{*}{Animal} 
    & \citet{li2019mixed} & Target registration error, Target deviation error (needle)\\
    & \citet{liu2021wearable} & Tracking accuracy \\
    & \citet{li2021augmented} & Target deviation error (marker), Task completion time, complications \\
\midrule          
\multirow{5}[2]{*}{Cadaver} & \citet{mcjunkin2018development} & Target registration error \\
    & \citet{muller2020augmented} & Target deviation error (k-wires)\\
    & \citet{tian2020validation} & Target registration error, Target deviation error \\
    & \citet{meng2021feasibility} & Target deviation error (osteotomy lines)\\
    & \citet{spirig2021augmented} & Target deviation error (k-wires) \\
\midrule         
Human & \citet{buch2021development} & Quant: Target registration error \\
\bottomrule
\toprule
\multicolumn{3}{c}{Studies in a Relevant Environment (Quantitative and qualitative measures)} \\
\midrule
\multirow{48}[2]{*}{\begin{sideways} Case studies \end{sideways}} 
    & \citet{aruanno2017hololens} & Quant: Number of successful completions \\
    & \citet{kuhlemann2017towards} & Quant: Target registration error; Qual: Likert questionnaire \\
    & \citet{perkins2017mixed} & Quant: Target visualization error \\
    & \citet{qian2017comparison} & Quant: Latency; Qual: Image quality, NASA-TLX \\
    & \citet{yamashita2017pedestrian} & Quant: Number of successful completions \\
    & \citet{condino2018build} & Quant: Target registration error; Qual: NASA TLX, Likert questionnaire \\
    & \citet{ingeson2018microsoft} & Qual: Likert questionnaire \\
    & \citet{jang2018three} & Quant: CPU usage, GPU usage, power usage, FPS; Qual: Likert questionnaire \\
    & \citet{pepe2018pattern} & Quant: Target visualization error; Qual: Likert questionnaire \\
    & \citet{sirilak2018new} & Qual: SUS, Likert questionnaire \\
    & \citet{amini2019augmented} & Quant: Task completion time; Qual: Likert questionnaire, SUS \\
    & \citet{aruanno2019memholo} & Quant: Number of successful completions, performance metrics; Qual: Likert questionnaire \\
    & \citet{balian2019feasibility} & Quant: Performance metrics; Qual: Likert questionnaire \\
    & \citet{blusi2019feasibility} & Qual: Likert questionnaire \\
    & \citet{brun2019mixed} & Qual: Likert questionnaires, surgical plan \\
    & \citet{checcucci20193d} & Qual: Likert questionnaires, surgical plan \\
    & \citet{gao2019feasibility} & Quant: Target deviation error (pointer), Task completion time \\
    & \citet{moosburner2019real} & Qual: SUS, Likert questionnaire \\
    & \citet{pepe2019marker} & Quant: Target visualization error; Qual: Likert questionnaire \\
    & \citet{qian2019aramis} & Quant: Target visualization error, latency, Task completion time, Number of successful completions \\
    & \citet{rose2019development} & Quant: Target registration error, Task completion time; Qual: Likert questionnaire \\
    & \citet{sun2019validity} & Quant: Measurement accuracy, Task completion time \\
    & \citet{bulliard2020preliminary} & Quant: Task completion time, number of interactions \\
    & \citet{cartucho2020multimodal} & Qual: Likert questionnaires \\
    & \citet{fischer2020evaluation} & Quant: Target registration error, Registration time; Qual: questionnaire \\
    & \citet{guerrero2020holonote} & Qual: Likert questionnaire \\
    & \citet{hong2020exploring} & Quant: Performance metrics \\
    & \citet{kumar2020use} & Qual: Likert questionnaire \\
    & \citet{koop2020hololens} & Quant: Measurement accuracy \\
    & \citet{muangpoon2020augmented} & Quant: Fiducial registration error; Qual: Likert questionnaire \\
    & \citet{nguyen2020baugmented} & Qual: preferences questionnaire \\
    & \citet{nuri2020augmented} & Quant: Target visualization error, Registration time \\
    \shrinkheight{-11\normalbaselineskip}
    & \citet{pelanis2020use} & Quant: Task completion time, Number of successful completions \\
    & \citet{rositi2021presentation} & Qual: Likert questionnaire \\
    & \citet{scherl2021augmented} & Quant: Target registration error; Qual: Likert questionnaire \\
    & \citet{southworth2020performance} & Quant: FPS, power usage, Latency; Qual: Image quality \\
    & \citet{zuo2020novel} & Quant: Target registration error; Qual: NASA TLX \\
    & \citet{bogomolova2021development} & Qual: Likert questionnaire \\
    & \citet{brunzini2021mixed} & Qual: Likert questionnaire, questionnaire;  \\
    & \citet{cofano2021augmented} & Qual: SUS \\
    & \citet{dennler2021augmented} & Qual: Likert questionnaire \\
    & \citet{gsaxner2021augmented} & Quant: Usage times; Qual: Likert questionnaire, SUS \\
    & \citet{heinrich2021holopointer} & Quant: Task completion time, performance; Qual: Likert questionnaire \\
    & \citet{koyachi2021accuracy} & Quant: Target deviation error (osteotomy lines) \\
    & \citet{kumar2021novel} & Qual: Likert questionnaire \\
    & \citet{schneider2021augmented} & Quant: Number of successful completions, Target deviation error (drain)\\
    & \citet{velazco2021evaluation} & Quant: Task completion time, Target deviation error (needles), interactions \\
\midrule       
\multirow{41}[2]{*}{\begin{sideways}Comparative Study\end{sideways}} 
    & \citet{wang2017augmented} & Quant: Task completion time; Qual: Likert questionnaire, questionnaire \\
    & \citet{andress2018fly} & Quant: Marker tracking accuracy, Target registration error, Target deviation error (k-wires), Number of X-Ray acquisitions \\
    & \citet{deib2018image} & Quant: Task completion time, Dosimetry \\
    & \citet{incekara2018clinical} & Quant: Target visualization error \\
    & \citet{sharma2018mixed} & Quant: Task completion time, Number of successful completions \\
    & \citet{angelopoulos2019enhanced} & Quant: Task completion time, performance metrics \\
    & \citet{li2019mixed} & Quant: Registration time, Target deviation error (drain)\\
    & \citet{palermo2019augmented} & Quant: Task completion time, Number of successful completions \\
    & \citet{rohrbach2019augmented} & Quant: Task completion time, performance metrics \\
    & \citet{stojanovska2019mixed} & Quant: Exam scores, Learning time \\
    & \citet{talaat2019three} & Quant: Distance between landmarks \\
    & \citet{van2019clinical} & Quant: Fiducial registration error \\
    & \citet{wake2019patient} & Qual: Likert questionnaire \\
    & \citet{al2020effectiveness} & Quant: Task completion time; Qual: Performance evaluation, Likert questionnaire \\
    & \citet{antoniou2020biosensor} & Quant: Biosignals \\
    & \citet{ferraguti2020augmented} & Quant: Registration accuracy, Task completion time, Target deviation error (needles); Qual: NASA-TLX \\
    & \citet{galati2020experimental} & Quant: Task completion time; Qual: Likert questionnaires \\
    & \citet{geerse2020quantifying} & Quant: Measurement accuracy \\
    & \citet{glick2020augmenting} & Quant: Task completion time; Qual: Likert questionnaires \\
    & \citet{gnanasegaram2020evaluating} & Quant: Exam scores, Qual: Likert questionnaire \\
    & \citet{house2020use} & Qual: Likert questionnaire \\
    & \citet{janssen2020effects} & Quant: Performance metrics; Qual: Likert questionnaire \\
    & \citet{qian2020flexivision} & Quant: Task completion time, Number of successful completions; Qual: NASA-TLX \\
    & \citet{robinson2020evaluating} & Quant: Exam scores; Qual: Self assessment \\
    & \citet{ruthberg2020mixed} & Quant: Exam scores, Learning time \\
    & \citet{ruger2020ultrasound} & Quant: Task completion time, Target deviation error (needles); Qual: NASA-TLX, questionnaire \\
    & \citet{saito2020intraoperative} & Qual: NASA-TLX \\
    & \citet{schoeb2020mixed} & Quant: Exam scores Qual: Self assessment, Likert questionnaires \\
    & \citet{condino2021hybrid} & Qual: Likert questionnaire, planning \\
    & \citet{gehrsitz2021cinematic} & Quant: Task completion time; Qual: Likert questionnaire \\
    & \citet{glas2021augmented} & Quant: Task completion time, Target deviation error (pointer); Qual: Questionnaire \\
    \shrinkheight{-10\normalbaselineskip}
    & \citet{dennler2021baugmented} & Quant: Target deviation error (screws), Number of successful completions \\
    & \citet{iqbal2021augmented} & Quant: Task completion time; Qual: Likert questionnaire \\
    & \citet{ivan2021augmented} & Quant: Target visualization error; Qual: Performance evaluation, \\
    & \citet{long2021comparison} & Quant: Target deviation error (needle), Task completion time, radiation \\
    & \citet{moro2021hololens} & Quant: Exam scores; Qual: Likert questionnaire \\
    & \citet{nguyen2022holous} & Quant: FPS, Latency, marker tracking accuracy, Task completion time \\
    \shrinkheight{-5\normalbaselineskip}
    & \citet{putnam2021virtual} & Qual: Likert questionnaire \\
    & \citet{qi2021holographic} & Quant: Registration time, Target visualization error, Task completion time \\
    & \citet{scherl2021baugmented} & Quant: Target registration error, Task completion time, number of complicaitons \\
    & \citet{suzuki2021learning} & Quant: Task completion time, Performance metrics; Qual: Questionnaire \\
    & \citet{van2021effect} & Quant: Target deviation error (drain); Qual: Performance evaluation, Likert questionnaire \\
\bottomrule
\end{xtabular*}
\vspace{7pt}
\twocolumn
}

\else
  \input{tables/eval_table_xtab}%
\fi
\makeatother


\section{Conclusion and outlook}
With \numpapers~original, peer reviewed works in the medical field, the HoloLens certainly had a large impact on medical AR already. In this systematic review, we found that, while various medical specialties and applications have been investigated, and a fair number of systems have been studied clinically, only few works have clinically demonstrated clear advantages of HoloLens-based systems over the current state-of-the-art. The acceptance of new technologies, such as AR, in the medical field is an ongoing challenge for researchers, medical professionals and patients alike. In this review, we identify that increased efforts in the areas of precision, reliability, usability, workflow and perception are necessary to establish AR in clinical practice. 

We found that applications targeted at physicians and healthcare professionals are, by far, the most common. While the potential benefit for AR supported image guidance and navigation is very high, those systems are also difficult to implement, mostly due to the high accuracy and reliability demands. The reviewed studies suggest that, for high precision applications, registration and tracking errors achieved with the HoloLens are generally too high, regardless of the employed technical paradigm and method. However, for procedures carried out without image guidance, for which sub-millimeter precision is not necessary (e.g., ablations~\cite{ferraguti2020augmented},  ventriculostomy~\cite{li2018wearable,van2021effect,schneider2021augmented} or certain orthopedic interventions~\cite{dennler2021baugmented}), the HoloLens is already a very promising tool. In these scenarios, the slim form factor and low cost of the HoloLens in comparison to traditional image guidance systems could make navigation feasible for procedures which have not benefited from it before. For this purpose, however, automatic and accurate inside-out registration and tracking is paramount to keep the setup workflow.

The second most common intended user group were students. In educational training scenarios, the HoloLens was shown to be an effective enhancement for medical simulators~\cite{balian2019feasibility, hong2020exploring, muangpoon2020augmented, schoeb2020mixed, suzuki2021learning, heinrich2021holopointer}, in particular for providing visual feedback during training tasks. In anatomy learning, the effects of HoloLens learning compared to conventional learning using cadavers or other computerized methods seem to be small, although several studies report improved engagement and motivation of students, which could have positive effects in the long term. Anatomy learning studies in this review also usually feature relatively simple, conventional 3D models. More innovative visualizations, including interactive, dynamic content, which can not be easily delivered by regular computerized methods, have not been explored in depth yet.

Finally, patients were the least frequent target user. Unfortunately, many interesting assistance and monitoring applications are limited by the restricted possible usage time of the HoloLens, and it is not foreseeable that next-gen OST-HMD devices will overcome these limitations in the near future. However, for selected scenarios, such as guidance through therapy and diagnosis sessions, the HoloLens has already shown to be useful. Until now, only a small number of applications have been explored in this context -- it will be interesting to see whether other disciplines can benefit from such paradigms as well.

The majority of studies in this review seeks a registration between real and virtual environment, and inside-out approaches, in particular using image fiducials, are the most common methods to achieve the required registration. This observation is unsurprising -- after all, such approaches are relatively easy to implement. Our analysis shows that they deliver a reliable, acceptable accuracy in controlled settings. Their described disadvantages, such as line-of-sight constraints and susceptibility to different viewing positions, movement patterns and lighting conditions, however, likely impede clinical adoption. Spherical markers seem to be more robust and encouraging results have been reported~\cite{kunz2020infrared}, more recently also for the HoloLens 2~\cite{gsaxnerVRST21}. Innovative, marker-less, inside-out strategies have been reported for registration, but are still hampered by technical limitations. For instrument tracking, research in the direction of marker-less, inside-out methods based on deep learning is only recently gaining traction~\cite{doughty2022hmd}, but will surely have a large impact in the field. 


When it comes to data and visualization, the majority of studies display pre-interventionally acquired 3D medical imaging data, primarily from CT or CTA, visualized through surface rendering. We expect this trend to continue. Compared to volume rendering, surface renderings are easy to create, modify and efficient to render, and no clear advantage of volume rendering through the HoloLens has been shown so far. Perceptual issues, in particular depth perception, are a known problem in AR~\cite{livingston2013pursuit}, and several works mention that incorrect depth perception negatively influenced the perceived accuracy of their application and impaired guidance through the HoloLens. Still, very few reports concern themselves with visualization strategies overcoming these limitations, and use very simple methods (e.g., wire frames~\cite{fischer2020evaluation}). While many strategies exist to improve depth perception in medical AR~\cite{gsaxner2021archapter}, most of them are difficult to apply with OST displays, such as the HoloLens, where only additive visual information is possible and the view of reality cannot be altered. Novel, innovative strategies will be necessary to overcome this limitation in the future. 

It is paramount that medical AR applications are validated with the intended user in the loop, and it is encouraging to see that the majority of studies in this review evaluate their applications in a relevant setting. Still, the large variety in experimental setups and acquired measures, together with the lack of standardized protocols, makes it very difficult to clinically validate these methods. We believe that this review can serve as a guideline to researchers, to help them in picking appropriate experimental protocols and measures for their scenario. We think that it is time for medical AR to step out of the comfort zone of controlled laboratory settings, and finally find its way into medical routine. To this end, close collaborations between researchers, universities, clinicians and patients, as well as comparative studies on a larger scale are necessary. 

The HoloLens 1 has likely reached the end of its life cycle in research, due to the release of its direct successor, but it has caused a major boost in medical AR research. With the availability of novel hardware, such as the HoloLens 2 or the Magic Leap 2, and the recent increased interest of other leading tech companies in AR technologies, we expect this trend to continue. Furthermore, specialized medical OST-HMD devices, e.g., xvision (Augmedics Inc., Arlington Heights, IL) or VOSTARS (University of Pisa, Pisa, IT), have the potential to address technical limitations in current, commercial devices. Improved hardware can also facilitate the use of deep learning models on the HMD itself, opening up countless possibilities in terms of recognition, tracking and scene understanding. In conclusion, we think that, although the feasibility of using the HoloLens for various medical scenarios has been suggested, research in medical AR is still in its early stages, and abundant areas for future work remain.


\section*{Acknowledgments}
This work received funding from the Austrian Science Fund (FWF) KLI 678: \lq enFaced - Virtual and Augmented Reality Training and Navigation Module for 3D-Printed Facial Defect Reconstructions', KLI 1044: \lq enFaced 2.0 - Instant AR Tool for Maxillofacial Surgery' and \lq KITE' (Plattform für KI-Translation Essen) from the REACT-EU initiative.

\printbibliography
\end{document}